\documentstyle[11pt,epsfig]{article}

\def\appendix{\par
 \setcounter{section}{0}
 \setcounter{subsection}{0}
 \def\thesection{Appendix \Alph{section}}
 \def\theequation{\Alph{section}.\arabic{equation}}
 \setcounter{equation}{0}}

\def\dis{\displaystyle}
\def\le{\left(}
\def\ri{\right)}

\def\no{\nonumber}

\def\e{\epsilon}
\def\f12{\frac{1}{2}}

\def\pd{\partial}

\def\L{\lambda}

\begin{document}
\begin{titlepage}
\flushright{USM-TH-209}
\vskip 2cm
\begin{center}
{\Large \bf Further results for the two-loop $Lcc$ vertex in the Landau gauge}\\
\vskip 1cm  
Gorazd Cveti\v{c}$^{a}$ and  Igor Kondrashuk$^{a,b}$\\
\vskip 5mm  
{\it  (a) Centro de Estudios Subat\'omicos y Departamento de F\'\i sica, \\
Universidad T\'ecnica Federico Santa Mar\'\i a, Casilla 110-V, Valpara\'\i so, Chile} \\
{\it  (b) Departamento de Ciencias B\'asicas, \\
Universidad del B\'\i o-B\'\i o, Campus Fernando May, Casilla 447, Chill\'an, Chile} \\
\end{center}
\vskip 20mm
\begin{abstract}
In the previous paper hep-th/0604112 we calculated the first of the five planar two-loop diagrams
for the $Lcc$ vertex of the general non-Abelian Yang-Mills theory, the vertex which allows us in
principle to obtain all other vertices via the Slavnov-Taylor identity. The integrand of
this first diagram has a simple Lorentz structure. In this letter we present the result 
for the second diagram, whose integrand has a complicated Lorentz structure. 
The calculation is performed in the $D$-dimensional Euclidean position space. 
We initially perform one of the two integrations in the position space and then reduce 
the Lorentz structure to $D$-dimensional scalar single integrals. Some of the latter 
are then calculated by the uniqueness method, others by the Gegenbauer polynomial technique. 
The result is independent of the ultraviolet and the infrared scale. It is expressed in 
terms of the squares of spacetime intervals between points of the effective fields in 
the position space -- it includes simple powers of these intervals, as well as logarithms 
and polylogarithms thereof, with some of the latter appearing within the Davydychev integral $J(1,1,1).$
Concerning the rest of diagrams, we present the result for the contributions correponding to third, fourth and fifth 
diagrams without giving the details of calculation. The full result for the $Lcc$ correlator of the effective 
action at the planar two-loop level is written explicitly for maximally supersymmetric Yang-Mills theory.  
\end{abstract}
\end{titlepage}

\section{Introduction}
\label{sec:intr}

It has been shown in Refs.~\cite{Cvetic:2004kx,Kondrashuk:2004pu,Cvetic:2006kk}
that the effective action of dressed mean fields for ${\cal N} =4$
super-Yang-Mills theory does not depend on any scale, ultraviolet or infrared.
These results were derived from the results of
Refs.~\cite{Cvetic:2002dx,Cvetic:2002in,Kondrashuk:2000br,Kondrashuk:2000qb,Kondrashuk:2003tw}.
Scale independence suggests that kernels of these dressed mean
fields can be analyzed  by the methods of conformal field theory.
We started to investigate the simplest scalar vertex $Lcc$ in the Landau gauge in
Ref.~\cite{Cvetic:2006iu}, where $c$ is a real ghost field and $L$ is the auxiliary
field which couples at the tree level to the BRST \cite{Becchi:1974md,Tyutin:1975qk}
transformation of the $c$ field. This vertex is simple in the Landau gauge where it is totally
finite for ${\cal N} =4$ super-Yang-Mills theory. By solving Slavnov-Taylor (ST) identity
\cite{Slavnov:1972fg,Taylor:1971ff,Slavnov:1974dg,Faddeev:1980be,Lee:1973hb,Zinn-Justin:1974mc},
all other vertices in that theory can be derived from this vertex \cite{Cvetic:2002dx}. The ST identity is a
consequence of the BRST symmetry of the classical action \cite{Becchi:1974md,Tyutin:1975qk}.
Recently, by using unitarity methods it has been demonstrated up to the four-loop level that
the only off-shell conformal integrals in the momentum space contribute 
in the maximally-helicity-violating  (MHV) four-particle amplitudes \cite{Bern:2006ew},
and iterative structure has been conjectured for all the MHV amplitudes \cite{Bern:2005iz}.

As has been found in Ref. \cite{Drummond:2006rz}, all the contributions to the off-shell 
four-point correlator of gluons in that theory up to three-loop level (at least) are equivalent to ladder integrals 
(UD integrals) calculated in the momentum space by Usyukina and Davydychev in Refs. \cite{Usyukina:1992jd,Usyukina:1993ch}.
They are expressed in terms of certain functions (UD functions).
It is known from Refs. \cite{Usyukina:1992jd,Usyukina:1993ch,Broadhurst:1993ib,Drummond:2006rz}  that UD functions are conformally invariant in the momentum space.
Fourier transform of UD functions in the momentum space can be expressed in terms of the 
UD functions in the position space which possess the same property 
of the conformal invariance but this time with the arguments of the position space. On the other hand,
the scale independence of the effective action of the dressed mean fields is a consequence of the 
vanishing of the beta function in the maximally supersymmetric Yang-Mills theory and the vanishing in turn is a 
consequence of the conformal symmetry of the theory. All this suggests that correlators of the 
dressed mean fields can be analyzed by the technique of the conformal field theory.

To solve the ST identity and to check the conformal invariance 
of the effective action of the dressed mean fields
explicitly, we should work in the position space. The two-loop contribution contains five diagrams.
We calculated the first contribution in the previous paper \cite{Cvetic:2006iu}.
Now we calculate the second contribution that corresponds to the diagram $(b)$ of Ref.~\cite{Cvetic:2006iu},
give the result for the rest of the diagrams, and present the full result for the planar two-loop 
contribution to the $Lcc$ correlator of the effective action in the Landau gauge for ${\cal N} =4$
supersymmetric Yang-Mills theory.  The notation used here is the same as in Ref.~\cite{Cvetic:2006iu}.
The $Lcc$ vertex is superficially convergent in the Landau gauge.
This fact can be checked by index counting and by noting that two derivatives
from the ghost propagators can always be integrated out of the diagram due to
the transversality of the gauge propagator. It means that the field $c$ does not have
renormalization in the Landau gauge. Formally, this result holds to all orders of
perturbation theory due to the so called antighost equation \cite{Blasi:1990xz}.
In the nonsupersymmetric theories this vertex is not finite and a calculation of the
anomalous dimension of operator $cc$ has been performed in \cite{Dudal:2003pe,Dudal:2003np}.  
Finiteness of this $Lcc$ vertex at planar two loop level in Landau gauge in maximally supersymmetric Yang-Mills theory 
is equivalent to the  cancellation of poles   between fourth and fifth diagrams of Ref.~\cite{Cvetic:2006iu} that is, in turn is 
equivalent to the vanishing of the $\beta$ function of the gauge coupling in that theory.

Knowing the structure of the $Lcc$ vertex one can obtain the structure of other 
irreducible vertices by solving the ST identity. The algorithm can be applied also
to other theories different from the theory in consideration. Among possible applications 
are ${\cal N}=8$ supergravity, Chern-Simons theory, string field theory, 
massless gauge theory near fixed points in the coupling space, and topological field 
theories in higher dimensions. 

The paper has the following structure. In Sec.~\ref{sec:diagb} we write the integral
expression that corresponds to the diagram $(b)$, and analyze it by dividing it 
in three parts. In Sec.~\ref{sec:red}, the result for the first part is reduced to
single $D$-dimensional integrals ($D = 4 - 2 \e$).
The analogous expressions for the second and the third part are written 
in \ref{App:A}. The sum of all three parts contains single $D$-dimensional
integrals $J(\alpha_1,\alpha_2,\alpha_3)$
with the sum of indices $\alpha_1+\alpha_2+\alpha_3$ in the denominators equal to
$D-1,$ $D,$ $D+1,$  $D+2.$
In Sec.~\ref{sec:D-1} we present the part with the integrals whose sum of indices is
equal to $D-1$. In \ref{App:B} we present the part with the integrals whose sum
of the indices is $D$, $D+1$,  and $D+2$; these integrals are
calculated there explicitly by using the 
uniqueness method and its variants \cite{Vasiliev:1981dg,Kazakov:1984bw}.
In Sec.~\ref{sec:D-1} we reduce the aforementioned $(D-1)$-type partial sum to 
explicitly known terms and to terms containing the 
integrals $J(1,1,1)$, $J(\e,2-3\e,1)$ and 
derivatives thereof -- by applying formulas obtained in \ref{App:C} 
where the method of integration by parts procedure (IBP) 
\cite{Tkachov:1981wb,Chetyrkin:1981qh} was used. 
In Sec.~\ref{sec:sum} we obtain the entire result for the
diagram $(b)$ in $D$ dimensions, in known terms and terms with 
the integrals  $J(1,1,1)$, $J(\e,2-3\e,1)$ and derivatives thereof.
The terms with $J(\e,2-3\e,1)$ are calculated explicitly in Sec.~\ref{sec:sum},
while derivatives of $J(\e,2-3\e,1)$ are calculated in \ref{App:D} --
in both cases those formulas of \ref{App:C} were employed which were
obtained by the Gegenbauer polynomial technique (GPT) introduced 
originally in Refs.~\cite{Chetyrkin:1980pr,Celmaster:1980ji,Terrano:1980af,Lampe:1982av}
and further developed in Ref.~\cite{Kotikov:1995cw}. 
In \ref{App:E}, all the contributions for the diagram $(b)$ in $D$ dimensions
are collected, and the $\e \to 0$ limit is performed. In Sec.~\ref{sec:finres}
we present this result in a more explicit and shorter form. In Sec.~7 
we write the result of calculation of the rest of diagram and present the 
total result for the planar two-loop $Lcc$ correlator of the effective action 
in the Landau gauge. In Conclusions, we comment on the result obtained and 
on further possible developments of the analysis performed in this paper. 
  
\section{Diagram $(b)$}
\label{sec:diagb}

\begin{figure}[ht]
\begin{center}
\epsfig{file=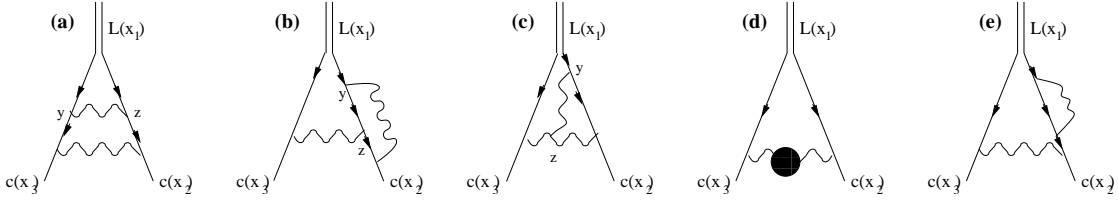, width=15.cm}
\end{center}
\vspace{0.0cm}
\caption{\footnotesize
The two-loop diagrams for the $Lcc$ vertex. The wavy lines represent 
the gluons, the straight lines the ghosts. 
Black disk stands for the total one-loop correction to the gluon propagator.}
\label{Lccfig}
\end{figure}

The two-loop correction to the $Lcc$ vertex can be represented as a sum of five 
diagrams depicted in Fig.~\ref{Lccfig}. 
We have introduced for brevity the notation 
$$[yz] = (y-z)^2, ~~~~ [y1] = (y-x_1)^2,....$$  
and so on.
The structure of Lorentz indices in diagram $(b)$ is more complicated than the 
structure of the indices in the diagram $(a)$ considered in Ref.~\cite{Cvetic:2006iu}.
The algebra can be inferred immediately from the diagram $(b)$.  
\begin{eqnarray*}
\dis{\frac{(2z)_\sigma}{[2z]^{2-\e}}\le \frac{g_{\sigma\lambda}}{[2y]^{1-\e}} 
+ 2(1-\e)\frac{(2y)_\sigma (2y)_\L}{[2y]^{2-\e}}\ri\frac{(y1)_\L}{[y1]^{2-\e}}
\frac{(zy)_\mu}{[zy]^{2-\e}}\le \frac{g_{\mu\nu}}{[3z]^{1-\e}} 
+ 2(1-\e)\frac{(3z)_\mu (3z)_\nu}{[3z]^{2-\e}}\ri\frac{(31)_\nu}{[31]^{2-\e}}}
\end{eqnarray*} 
We work in the position space, in $D=4 - 2 \e$ dimensions, using the technique of
dimensional regularization \cite{Bollini:1972ui}, thus maintaining the intermediate
results finite.
We have to calculate the $2 D$-dimensional integral\footnote{
We use the $D$-dimensional measure $Dx \equiv  \pi^{-\frac{D}{2}}d^Dx$   \cite{Cvetic:2006iu}.} 
\begin{eqnarray}
\int~Dy~Dz~\frac{(2z)_\sigma}{[2z]^{2-\e}}\le \frac{g_{\sigma\lambda}}{[2y]^{1-\e}} 
+ 2(1-\e)\frac{(2y)_\sigma (2y)_\L}{[2y]^{2-\e}}\ri\frac{(y1)_\L}{[y1]^{2-\e}}\times\no\\
\times\frac{(zy)_\mu}{[zy]^{2-\e}}\le \frac{g_{\mu\nu}}{[3z]^{1-\e}} 
+ 2(1-\e)\frac{(3z)_\mu (3z)_\nu}{[3z]^{2-\e}}\ri\frac{(31)_\nu}{[31]^{2-\e}} \ .
\label{start0}
\end{eqnarray} 
By using the representation $(y1)_\L = (y2)_\L + (21)_\L$, after simple algebra we can
represent this integral as a sum of three integrals
\begin{eqnarray*}
-(3-2\e)\int~Dy~Dz~\frac{(2z)_\sigma(2y)_\sigma}{[y1]^{2-\e}[2z]^{2-\e}[2y]^{1-\e}}
\frac{(zy)_\mu}{[zy]^{2-\e}}\le \frac{g_{\mu\nu}}{[3z]^{1-\e}} + 
2(1-\e)\frac{(3z)_\mu (3z)_\nu}{[3z]^{2-\e}}\ri\frac{(31)_\nu}{[31]^{2-\e}} \\
+(2-2\e)\int~Dy~Dz~\frac{[(2z)_\sigma(2y)_\sigma][(2y_\L)(21)_\L]}
{[y1]^{2-\e}[2z]^{2-\e}[2y]^{1-\e}}
\frac{(zy)_\mu}{[zy]^{2-\e}}\le \frac{g_{\mu\nu}}{[3z]^{1-\e}} + 
2(1-\e)\frac{(3z)_\mu (3z)_\nu}{[3z]^{2-\e}}\ri\frac{(31)_\nu}{[31]^{2-\e}} \\
+ \int~Dy~Dz~\frac{(2z)_\sigma(21)_\sigma}{[y1]^{2-\e}[2z]^{2-\e}[2y]^{1-\e}}
\frac{(zy)_\mu}{[zy]^{2-\e}}\le \frac{g_{\mu\nu}}{[3z]^{1-\e}} + 
2(1-\e)\frac{(3z)_\mu (3z)_\nu}{[3z]^{2-\e}}\ri\frac{(31)_\nu}{[31]^{2-\e}}.
\end{eqnarray*} 
As in the previous paper \cite{Cvetic:2006iu}, we modify the $\e$'s appearing
in the propagators ($\e \mapsto \kappa_j \e$) in order to make the uniqueness method
applicable and thus to make the calculations easier. 
Such changes do not affect our result in the $\e \to 0$ limit since the integral is
finite. We do the following changes: 
in the powers of the denominators of the four ghost propagators in Eq.~(\ref{start0}), 
we change $\e \mapsto 2 \e$ in the first one, and $\e \mapsto 0$
in the other three;
in the powers of the denominators in the two gauge propagators, 
we change $\e \mapsto 2 \e$ in the first
and $\e \mapsto 0$ in the second propagator;
in the nominator of the second gauge propagator, we change $\e \mapsto \bar\e$,
where $\bar\e$ is as in the previous paper;
in the nominator of the first gauge propagator, we change $\e \mapsto \bar\e_ 2$ 
where $\bar\e_2$ is different because it is related to the requirement of keeping 
the transversality in the modified propagators \cite{Cvetic:2006iu}
\begin{equation}
\bar\e = - \frac{2 \e}{(1 - 2 \e)} \ , \qquad
\bar\e_2 = \frac{4 \e}{(1  + 2 \e)} \ .
\label{eps}
\end{equation}
The above integral for the diagram $(b)$ now acquires the following form:  
\begin{eqnarray}
I_b &=& 
\int~Dy~Dz~\frac{(2z)_\sigma}{[2z]^{2-2\e}}\le \frac{g_{\sigma\lambda}}{[2y]^{1-2\e}} 
+ 2(1-\bar{\e}_2)\frac{(2y)_\sigma (2y)_\L}
{[2y]^{2-2\e}}\ri\frac{(y1)_\L}{[y1]^{2}}
\no\\
&& \times
\frac{(zy)_\mu}{[zy]^2}\le \frac{g_{\mu\nu}}{[3z]} + 
2(1-\bar{\e})\frac{(3z)_\mu (3z)_\nu}{[3z]^2}\ri\frac{(31)_\nu}{[31]^2} 
\no\\
& = & - (3 - 2 \bar\e_2) I_{b,1} + (2 - 2 \bar\e_2) I_{b,2} + I_{b,3} \ ,
\label{start}
\end{eqnarray}
where the three integrals $I_{b,j}$ are
\begin{eqnarray}
I_{b,1} & = & \int~Dy~Dz~\frac{(2z)_\sigma(2y)_\sigma}{[y1]^{2}[2z]^{2-2\e}[2y]^{1-2\e}}
\frac{(zy)_\mu}{[zy]^2}\le \frac{g_{\mu\nu}}{[3z]} 
+ 2(1-\bar\e)\frac{(3z)_\mu (3z)_\nu}{[3z]^{2}}\ri\frac{(31)_\nu}{[31]^2} \ ,
\label{Ib1def}
\\
I_{b,2} & = & \int~Dy~Dz~\frac{[(2z)_\sigma(2y)_\sigma][(2y_\L)(21)_\L]}
{[y1]^{2}[2z]^{2-2\e}[2y]^{2-2\e}}
\frac{(zy)_\mu}{[zy]^2}\le \frac{g_{\mu\nu}}{[3z]} 
+ 2(1-\bar\e)\frac{(3z)_\mu (3z)_\nu}{[3z]^2}\ri\frac{(31)_\nu}{[31]^2} \ ,
\label{Ib2def}
\\
I_{b,3} & = & \int~Dy~Dz~\frac{(2z)_\sigma(21)_\sigma}{[y1]^{2}[2z]^{2-2\e}[2y]^{1-2\e}}
\frac{(zy)_\mu}{[zy]^2}\le \frac{g_{\mu\nu}}{[3z]} 
+ 2(1-\bar\e)\frac{(3z)_\mu (3z)_\nu}{[3z]^2}\ri\frac{(31)_\nu}{[31]^2} \ .
\label{Ib3def}
\end{eqnarray}
In these $2D$-dimensional integrals, we first perform the Lorentz algebra only
for certain subproducts in the numerators, then perform one of the two
$D$-dimensional integrations by using the uniqueness method or its variants,\footnote{
By the ``variants of the uniqueness method'' we mean the cases when the sum
of the powers in the denominator of the integrand is larger than $D$ but can be made equal $D$ once
the integrand is represented as a derivative, or derivatives, with respect to a coordinate that
is not integrated over.}
 and then finish the Lorentz algebra. 
We found this method simpler than making the entire Lorentz algebra initially and then 
integrating the contributions. After integrating, 
the number of terms in the algebra is significantly reduced.

\section{Reduction to single scalar integrals} \label{reduction}
\label{sec:red}

Following the afore-described procedure, we can scalarize the expression 
for each of the three $2D$-dimensional integrals in Eq.~(\ref{start}),
reducing them to linear combinations of integrals $J(\alpha_1,\alpha_2,\alpha_3)$
where
\begin{displaymath}
J(\alpha_1,\alpha_2,\alpha_3) = 
\int Dx \frac{1}{[x1]^{\alpha_1}[x2]^{\alpha_2}[x3]^{\alpha_3}}  \ .
\end{displaymath}  
For the first integral we obtain
\begin{eqnarray}
I_{b,1} =  
\frac{A(1,1-2\e,1)}{4\e(1-2\e)^2}\left\{\le-\frac{6(1-\e^2)(1-2\e)}{[31]^2[12]^{-\e}} 
- \frac{2(1-\e^2)(1-2\e)}{[31][12]^{1-\e}}\ri J(1+\e,2-3\e,1) \right.\no\\
+ \left. \frac{2(1-\e^2)(1-2\e)}{[31]^2[12]^{1-\e}} J(\e,2-3\e,1)  
+ \frac{2(1-\e^2)(1-2\e)}{[31][12]^{1-\e}} J(2+\e,1-3\e,1)  \right.  \no\\ 
+ \left. \frac{2(1-\e^2)(1-2\e)}{[31]^2[12]^{1-\e}} J(2+\e,1-3\e,0) 
-  \frac{(1-2\e^2)(1-2\e)}{[31]^2[12]^{1-\e}}J(1+\e,1-3\e,1)  \right. \no\\
+ \left. \frac{6(1-\e^2)(1-2\e)}{[31][12]^{-\e}}J(2+\e,2-3\e,1)  
- \frac{2(1-\e^2)(1-2\e)}{[31]^{2}[12]^{1-\e}}J(1+\e,2-3\e,0)  \right. \no\\
+  \left. \frac{\e - \e^2}{[31][23]^{1-\e}} J(2,1-3\e,1+\e) 
+  \frac{\e-\e^2}{[31]^2[23]^{1-\e}} J(1,2-3\e,\e)  \right. \no\\
-  \left. \frac{4-8\e+3\e^2}{[31][23]^{-\e}}J(2,2-3\e,1+\e) 
+ \frac{4-8\e+3\e^2}{[31]^2[23]^{-\e}}J(1,2-3\e,1+\e) \right.\no  \\
+ \left. \le\frac{\e(1-2\e)}{[31][23]^{1-\e}} - \frac{4-10\e+6\e^2}{[31]^2[23]^{-\e}} 
- \frac{[12]\e(2-3\e)}{[31]^2[23]^{1-\e}}\ri J(2,2-3\e,\e) \right.  \no\\ 
+ \left. \left[-\frac{(1-2\e)(1-\e)}{[31]^2[23]^{1-\e}} 
- \frac{(1-\e)^2}{[31][23]^{2-\e}}  
+ \frac{[12](1-\e)^2}{[31]^2[23]^{2-\e}} \right] J(2,1-3\e,\e)  \right. \no\\ 
+ \left. \left[-\frac{2\e(1-\e)}{[31]^2[23]^{1-\e}} 
- \frac{\e(1-\e)}{[31][23]^{2-\e}}  
+ \frac{[12]\e(1-\e)}{[31]^2[23]^{2-\e}} \right] J(2,2-3\e,\e-1) 
\right. \no\\  
+ \left. \frac{1-2\e}{[12]^{1-\e}[31]} J(1+\e,1-3\e,2)  
- \frac{1-2\e}{[12]^{1-\e}[31]^{2}} J(\e,1-3\e,2 ) \right. \no \\
+ \left. \le \frac{1-\e}{[31][23]^{2-\e}} + \frac{1-\e}{[31]^{2}[23]^{1-\e}}  
- \frac{[12](1-\e)}{[31]^{2}[23]^{2-\e}} \ri J(2,1-2\e,0) \right. \no\\
\left. - \frac{\e-\e^2}{[31]^{2}[23]^{1-\e}}J(1,1-3\e,1+\e)\right\} \ .
\label{idn}
\end{eqnarray}
Here we use the notation
\begin{displaymath}
A(\alpha_1,\alpha_2,\alpha_3) = a(\alpha_1)a(\alpha_2)a(\alpha_2); \qquad
a(\alpha) = \frac{\Gamma(D/2 - \alpha)}{ \Gamma(\alpha)} \ .
\end{displaymath}
The second and the third integrals (\ref{Ib2def}) and (\ref{Ib3def}) are given in
\ref{App:A}, Eqs.~(\ref{second}) and (\ref{third}), respectively,
and have a similar structure as the first integral (\ref{idn}). 
As a consequence, the total Integral $I_b$ can be organized into the sum
\begin{equation}
I_b = \sum_{k=D-1}^{D+2} I_b^{(k)} \ ,
\label{Ibsumind}
\end{equation}
where $I_b^{(k)}$ represents the terms with $J(\alpha_1,\alpha_2,\alpha_3)$ with
$\alpha_1 + \alpha_2 + \alpha_3 = k$.

\section{Integrals with the sum of indices $D-1$} 
\label{section(D-1)} \label{sec:D-1}

The result for $I_b^{(D)}$ and for $I_b^{(D+1)}+I_b^{(D+2)}$ is written
in \ref{App:B} in terms of $D$-dimensional $J$-integrals and 
then reduced to explicit expressions
by using the uniqueness relation and its variants for the integrals.
Below we consider the sum $I_b^{(D-1)}$.
\begin{eqnarray}
I_b^{(D-1)} = \frac{A(1,1-2\e,1)}{4\e(1+2\e)(1-2\e)^2} 
\left\{  \frac{-1+5\e-6\e^2}{[31]^2[12]^{1-\e}} J(\e,2-3\e,1) \right. \no\\  
+ \left.  \left[\frac{1-7\e+10\e^2}{[31]^2[23]^{1-\e}} 
+ \frac{1-4\e+5\e^2}{[31][23]^{2-\e}} 
- \frac{(1-4\e+5\e^2)[21]}{[31]^2[23]^{2-\e}}  \right] J(1,2-3\e,\e)  \right. \no\\
+ \left. \left[\frac{2-6\e+4\e^2}{[31]^2[23]^{1-\e}} 
+ \frac{2-4\e+2\e^2}{[31][23]^{2-\e}}  
- \frac{[12](2-4\e+2\e^2)}{[31]^2[23]^{2-\e}} \right] J(2,1-3\e,\e) \right.   \no\\ 
+ \left. \left[\frac{4\e-4\e^2}{[31]^2[23]^{1-\e}} + \frac{2\e-2\e^2}{[31][23]^{2-\e}}  
+ \frac{[12](-2\e+2\e^2)}{[31]^2[23]^{2-\e}} \right] J(2,2-3\e,\e-1)   \right.  \no\\  
+ \left. \frac{-\e+2\e^2}{[12]^{1-\e}[31]^{2}} J(\e,1-3\e,2 ) \right. \no\\
+ \left. \frac{1-\e-2\e^2}{[12]^{1-\e}[31]^2} J(\e-1,2-3\e,2)  \right. \no\\
+  \left. \frac{-2\e+4\e^2}{[12]^{1-\e}[31]^2} J(\e-1,3-3\e,1)  \right. \no\\
+ \left.  \le \frac{2\e-6\e^2}{[31]^2[23]^{1-\e}} + \frac{\e-3\e^2}{[31][23]^{2-\e}} 
+ \frac{(-\e+3\e^2)[12]}{[31]^2[23]^{2-\e}}  \ri J(1,3-3\e,\e-1)\right\} \label{idn2}
\ .
\end{eqnarray} 
The $D$-dimensional integrals appearing in $I_b^{(D-1)}$ cannot be found
by the uniqueness method or its variants. We will reduce them in this Section
to explicit expressions and to terms proportional to the integrals
$J(1,1,1)$, $J(\e,2 - 3 \e,1)$, $J(1,2 - 3 \e,\e)$ and specific derivatives thereof
-- by applying the integration by parts (IBP) procedure \cite{Tkachov:1981wb,Chetyrkin:1981qh}.
Using formula (\ref{aux1}) from \ref{App:C} (IBP), we represent integral (\ref{idn2}) as   
\begin{eqnarray*}
I_b^{(D-1)} =
\frac{A(1,1-2\e,1)}{4\e(1+2\e)(1-2\e)^2}
\left\{ \frac{-1+4\e-(9/2)\e^2}{[31]^2[12]^{1-\e}} J(\e,2-3\e,1) \right. \\  
+ \left.  \left[\frac{1-6\e+(15/2)\e^2}{[31]^2[23]^{1-\e}} 
+ \frac{1-7/2\e+15/4\e^2}{[31][23]^{2-\e}} 
+ \frac{(-1+7/2\e-15/4\e^2)[21]}{[31]^2[23]^{2-\e}}  \right] J(1,2-3\e,\e)  \right. \\
+ \left. \left[\frac{2-6\e+4\e^2}{[31]^2[23]^{1-\e}} 
+ \frac{2-4\e+2\e^2}{[31][23]^{2-\e}}  
- \frac{[12](2-4\e+2\e^2)}{[31]^2[23]^{2-\e}} \right] J(2,1-3\e,\e)  
 \right. \\
+ \left. \frac{-\e+2\e^2}{[12]^{1-\e}[31]^{2}} J(\e,1-3\e,2 )\right.   \no\\ 
+ \left. \left[\frac{3\e-(5/2)\e^2}{[31]^2[23]^{1-\e}} 
+ \frac{(3/2)\e-(5/4)\e^2}{[31][23]^{2-\e}}  
+ \frac{[12](-(3/2)\e+(5/4)\e^2)}{[31]^2[23]^{2-\e}} \right] J(2,2-3\e,\e-1)   \right.  \no\\  
+ \left. \frac{1-(5/2)\e^2}{[12]^{1-\e}[31]^2} J(\e-1,2-3\e,2) \right\}\\
+ \frac{A(1,1-2\e,1) A(1,1-3\e,1+\e)}{8\e^2(1+2\e)(1-2\e)^2}
\le \frac{-\e - (1/2)\e^2}{[12]^{2-2\e}[23]^{1-2\e}[31]^{1+2\e}} 
+ \frac{-(\e/2) - (1/4)\e^2}{[12]^{2-2\e}[23]^{2-2\e}[31]^{2\e}} \right.\\
+ \left.  \frac{(3/2)\e+(7/4)\e^2}{[12]^{1-2\e}[23]^{2-2\e}[31]^{1+2\e} }  \ri  
\ .
\end{eqnarray*} 
After applying formulas (\ref{aux3}) and (\ref{aux4}) (IBP), 
this formula takes the following form:
\begin{eqnarray*}
I_b^{(D-1)} =
\frac{A(1,1-2\e,1)}{4\e(1+2\e)(1-2\e)^2}
\left\{ \frac{-1+5\e-(19/2)\e^2}{[31]^2[12]^{1-\e}} J(\e,2-3\e,1) \right. \\  
+ \left. \left[
\frac{-1+6\e-(29/2)\e^2}{[31]^2[23]^{1-\e}} 
+ \frac{-1+13/2\e-41/4\e^2}{[31][23]^{2-\e}} 
+ \frac{(1-13/2\e+41/4\e^2)[21]}{[31]^2[23]^{2-\e}}  
\right] J(1,2-3\e,\e)  \right. \\
+ \left. \left[\frac{3\e-(5/2)\e^2}{[31]^2[23]^{1-\e}} 
+ \frac{(3/2)\e-(5/4)\e^2}{[31][23]^{2-\e}}  
+ \frac{[12](-(3/2)\e+(5/4)\e^2)}{[31]^2[23]^{2-\e}} \right] J(2,2-3\e,\e-1)   \right.  \no\\  
+ \left. \frac{1-(5/2)\e^2}{[12]^{1-\e}[31]^2} J(\e-1,2-3\e,2) \right\}\\
\\
- \frac{1}{2}\left[ \frac{1}{[31]^2[23]} + \frac{1}{[31][23]^2}  
- \frac{[12]}{[31]^2[23]^2} \right] J(1,1,1) \\
+ \frac{A(1,1-2\e,1)A(1,1-3\e,1+\e)}{8\e^2(1+2\e)(1-2\e)^2} 
\left[\frac{-2+7\e-6\e^2}{[12]^{1-2\e}[23]^{1-2\e}[31]^{2+2\e}} 
+ \frac{-2+(11/2)\e-(1/4)\e^2}{[12]^{1-2\e}[23]^{2-2\e}[31]^{1+2\e} } \right. \\
+ \left. \frac{(2-4\e+2\e^2)}{[12]^{-2\e}[23]^{2-2\e}[31]^{2+2\e}} 
+  \frac{-\e - (1/2)\e^2}{[12]^{2-2\e}[23]^{1-2\e}[31]^{1+2\e}} 
+ \frac{-(\e/2) - (1/4)\e^2}{[12]^{2-2\e}[23]^{2-2\e}[31]^{2\e}} \right] \ .
\end{eqnarray*} 
Finally, applying relations (\ref{aux5}) derived in \ref{App:C}, we obtain the representation 
\begin{eqnarray} 
I_b^{(D-1)} =
\frac{A(1,1-2\e,1)}{4\e(1+2\e)(1-2\e)^2}
\left\{ \frac{5\e-12\e^2}{[31]^2[12]^{1-\e}} J(\e,2-3\e,1) \right. \no\\  
+ \left. \left[\frac{-1+9\e-17\e^2}{[31]^2[23]^{1-\e}} 
+ \frac{-1+8\e-(23/2)\e^2}{[31][23]^{2-\e}} 
+ \frac{(1-8\e+(23/2)\e^2)[21]}{[31]^2[23]^{2-\e}}  \right] J(1,2-3\e,\e)  \right. \no\\
+ \left. \left[\frac{3\e-(5/2)\e^2}{[31]^2[23]^{1-\e}} 
+ \frac{-(3/2)\e+(5/4)\e^2}{[31][23]^{2-\e}}  
+ \frac{[12](-(3/2)\e+(5/4)\e^2)}{[31]^2[23]^{2-\e}} \right]
(13)_\mu \pd_{\mu}^{(1)}J(1,2-3\e,\e) \right.  \no\\  
+ \left. \frac{1-(5/2)\e^2}{[12]^{1-\e}[31]^2} 
(31)_\mu \pd_{\mu}^{(3)}J(\e,2-3\e,1) \right\}\no\\
\no\\
- \frac{1}{2}\left[ \frac{1}{[31]^2[23]} + \frac{1}{[31][23]^2}  
- \frac{[12]}{[31]^2[23]^2} \right] J(1,1,1) \no\\
+ \frac{A(1,1-2\e,1)A(1,1-3\e,1+\e)}{8\e^2(1+2\e)(1-2\e)^2} 
\left[\frac{-2+7\e-6\e^2}{[12]^{1-2\e}[23]^{1-2\e}[31]^{2+2\e}} 
+ \frac{-3+6\e-1/2\e^2}{[12]^{1-2\e}[23]^{2-2\e}[31]^{1+2\e} } \right. \no\\
+ \left. \frac{2-4\e+2\e^2}{[12]^{-2\e}[23]^{2-2\e}[31]^{2+2\e}} 
+  \frac{-4\e - \e^2}{[12]^{2-2\e}[23]^{1-2\e}[31]^{1+2\e}} 
+ \frac{-2\e - \e^2/2}{[12]^{2-2\e}[23]^{2-2\e}[31]^{2\e}} \right] \label{idn3} \ .
\end{eqnarray}

\section{The sum}
\label{sec:sum}

Summing up Eqs. (\ref{idn3}), (\ref{D}) and (\ref{D+1}), we obtain  
\begin{eqnarray} 
I_b =
\frac{A(1,1-2\e,1)}{4\e(1+2\e)(1-2\e)^2}
\left\{ \frac{5\e-12\e^2}{[31]^2[12]^{1-\e}} J(\e,2-3\e,1) \right. \no\\  
+ \left. \left[\frac{-1+9\e-17\e^2}{[31]^2[23]^{1-\e}} 
+ \frac{-1+8\e-(23/2)\e^2}{[31][23]^{2-\e}} 
+ \frac{(1-8\e+(23/2)\e^2)[21]}{[31]^2[23]^{2-\e}}  \right] J(1,2-3\e,\e)  \right. \no\\
+ \left. \left[\frac{3\e-(5/2)\e^2}{[31]^2[23]^{1-\e}} 
+ \frac{-(3/2)\e+(5/4)\e^2}{[31][23]^{2-\e}}  
+ \frac{[12](-(3/2)\e+(5/4)\e^2)}{[31]^2[23]^{2-\e}} 
\right](13)_\mu \pd_{\mu}^{(1)}J(1,2-3\e,\e) \right.  \no\\  
+ \left. \frac{1-(5/2)\e^2}{[12]^{1-\e}[31]^2} 
(31)_\mu \pd_{\mu}^{(3)}J(\e,2-3\e,1) \right\}\no\\
\no\\
- \frac{1}{2}\left[ \frac{1}{[31]^2[23]} + \frac{1}{[31][23]^2}  
- \frac{[12]}{[31]^2[23]^2} \right] J(1,1,1) \no\\
+ \frac{1}{8} \left\{\frac{-19/4}{[12][23][31]^{2}} + \frac{5}{[12][23]^{2}[31]} 
+  \frac{1/4}{[12]^{2}[23][31]} 
+ \frac{- 11/4}{[12]^{2}[23]^{2}} + \frac{- 9/4}{[23]^{2}[31]^{2}} 
+  \frac{5/2}{[12]^{2}[31]^{2}}  \right\}\no\\
+ \frac{A(1,1-2\e,1)A(1,1-3\e,1+\e)}{8\e(1+2\e)(1-2\e)^2} 
\left\{\frac{-7/2}{[12]^{1-2\e}[23]^{1-2\e}[31]^{2+2\e}}  
+ \frac{-2}{[12]^{1-2\e}[23]^{2-2\e}[31]^{1+2\e} } \right. \no\\
+ \left. \frac{-1/2}{[12]^{-2\e}[23]^{2-2\e}[31]^{2+2\e}} 
+  \frac{3/2}{[12]^{2-2\e}[23]^{1-2\e}[31]^{1+2\e}} 
+ \frac{3/2}{[12]^{2-2\e}[23]^{2-2\e}[31]^{2\e}} \right.\no\\
+ \left. \frac{3}{[12]^{2-2\e}[23]^{-2\e}[31]^{2+2\e}} 
+ \frac{-6}{[31]^2[12]^{2-2\e}} +  \frac{1}{[31]^2[23]^{2-2\e}} \right\}\no\\
\no\\
+ \frac{A^2(1,1-2\e,1)}{4\e^2(1+2\e)(1-2\e)^2}
\left\{\frac{1-\e}{[12]^{1-\e}[23]^{2-\e}[31]} 
+ \frac{1-\e}{[12]^{1-\e}[23]^{1-\e}[31]^2} - 
\frac{1 - \e}{[12]^{-\e}[23]^{2-\e}[31]^2} \right\} \ .
\label{idn4}
\end{eqnarray} 
After applying formulas (\ref{aux6}) which were obtained by an application of
the Gegenbauer polynomial technique (GPT, cf.~Ref.~\cite{Kotikov:1995cw}), 
we obtain for the terms proportional
to $J(\e,2 - 3 \e,1)$ and $J(1, 2 - 3 \e, \e)$ in $I_b$ the following explicit result:
\begin{eqnarray} 
I_b^{(J)} \equiv
\frac{A(1,1-2\e,1)}{4\e(1+2\e)(1-2\e)^2}\left\{ \frac{5\e-12\e^2}{[31]^2[12]^{1-\e}} 
J(\e,2-3\e,1) \right. \no\\  
+ \left. \left[\frac{-1+9\e-17\e^2}{[31]^2[23]^{1-\e}} 
+ \frac{-1+8\e-(23/2)\e^2}{[31][23]^{2-\e}} 
+ \frac{(1-8\e+(23/2)\e^2)[21]}{[31]^2[23]^{2-\e}}  \right] 
J(1,2-3\e,\e)  \right\} \no\\
= \frac{A(1,1-2\e,1) A(1- 3\e,1,1+\e)}{8\e(1+2\e)(1-2\e)^2} 
\left[\frac{11}{[12]^{1-\e}[31]^2[23]^{1-\e}} \right.\no\\
+ \left. \frac{5}{[12]^{1-\e}[31][23]^{2-\e}} 
+ \frac{-5}{[12]^{-\e}[31]^2[23]^{2-\e}} \right]  \no\\
+ \frac{A(1,1-2\e,1) A(1- 3\e,1,1+\e)}{8\e^2(1+2\e)(1-2\e)^2}
\left[\frac{-1}{[12]^{1-\e}[31]^2[23]^{1-\e}} \right.\no\\
+ \left. \frac{-1}{[12]^{1-\e}[31][23]^{2-\e}} 
+ \frac{1}{[12]^{-\e}[31]^2[23]^{2-\e}}  \right]  \no\\
+ \frac{A(1,1-2\e,1) A(1- 3\e,1,1+\e)}{8\e(1+2\e)(1-2\e)^2}
\left[\frac{1}{[12]^{1-\e}[31]^2[23]^{1-\e}} \right.\no\\ 
+ \left. \frac{1}{[12]^{1-\e}[31][23]^{2-\e}} 
- \frac{1}{[12]^{-\e}[31]^2[23]^{2-\e}}  \right]\ln\frac{[23]}{[12]} \no\\
\no\\
+ \frac{1}{8}\left[\frac{4}{[12][31]^2[23]} + \frac{7/2}{[12][31][23]^{2}} 
+ \frac{-(7/2)}{[31]^2[23]^{2}} \right] \no \\
- \frac{1}{8}\left[\frac{1}{[12][31]^2[23]} + \frac{5}{[12][31][23]^{2}} 
+ \frac{-5}{[31]^2[23]^{2}}  \right]\ln\frac{[23]}{[12]} \no\\
+ \frac{1}{16}\left[\frac{1}{[12][31]^2[23]} + \frac{1}{[12][31][23]^2} 
+ \frac{-1}{[31]^2[23]^{2}} \right]\ln\frac{[12][23]}{[13]^2}\ln\frac{[23]}{[12]}  \no\\
+ \frac{1}{8}\left[\frac{2}{[31]^2[23]} - \frac{[12]}{[31]^2[23]^{2}} 
+ \frac{1}{[12][23]^2}   - \frac{1}{[31]^2[12]} \right] J(1,1,1) \ .
\label{sum of first two lines}
\end{eqnarray} 
The explicit results for the terms $I_b^{(\partial J,1)}, I_b^{(\partial J,2)}$ 
proportional to derivatives of $J(1, 2 - 3 \e, \e)$ and
$J(\e,2 - 3 \e,1)$ in $I_b$ are written in \ref{App:D} in Eqs.~(\ref{thirdline}) and
(\ref{fourthline}), respectively.

\section{The final result for diagram $(b)$}
\label{sec:finres}

Substitution of expressions (\ref{sum of first two lines}), (\ref{thirdline}) and (\ref{fourthline})
into the sum (\ref{idn4}), and its expansion in terms of $\e$, is performed in
\ref{App:E}, Eq.~(\ref{2001}). Performing explicitly the derivatives appearing in Eq. (\ref{2001}), 
we obtain the final result for the diagram $(b)$:
\begin{eqnarray} 
I_b(\e \to 0) =
\frac{1}{8} \left[\frac{-3\ln [23]  - 4 \ln[12] +7\ln [13]}{[12][23][31]^2} 
+ \frac{ -8\ln [23]  + 4 \ln[12] + 4 \ln [13]}{[12][23]^2[31]} \right. \no\\
+ \left.  \frac{ 3\ln [23]  - 3 \ln [13]}{[12]^2[23][31]}   + \frac{ 5\ln [23]  - 6\ln[12] 
+ \ln [13]}{[23]^2[31]^2}    + \frac{ 6\ln [23]  - 6\ln [13]}{[12]^2[31]^2}  \right.\no\\
+ \left.  \frac{ 3\ln [23]  - 3\ln [13]}{[12]^2[23]^2}  \right]  \no\\
+ \frac{1}{4}\left[\frac{2[12]}{[31]^2[23]^{2}} + \frac{1}{[12][23]^2}   
+ \frac{-1}{[31]^2[12]}  + \frac{-3}{[31][23]^2}\right] J(1,1,1) \no\\
+ \frac{1}{8} \left\{\frac{-2}{[12][23][31]^{2}} + \frac{8}{[12][23]^{2}[31]}  
+ \frac{- 3}{[12]^{2}[23]^{2}} + \frac{-3}{[23]^{2}[31]^{2}} 
+  \frac{3}{[12]^{2}[31]^{2}}  \right\} \ .
\label{b-final}
\end{eqnarray} 
The result for the Davydychev integral $J(1,1,1)$ is in Ref.~\cite{Davydychev:1992xr}.  
A new integral representation for it has been found in Ref.~\cite{Cvetic:2006iu}. The term linear in $\e$ of the $J(1,1,1)$ integral 
in $D=4-2\e$ dimensions was obtained in Ref.  \cite{Davydychev:1995mq}, while all-order $\e$-expansion 
was derived in Ref.  \cite{Davydychev:2000na}.

\section{The final result for $Lcc$ correlator}

In this section we write the total result for planar two-loop $Lcc$ vertex. It contains contribution of  five planar diagrams  $(a)-(e).$ 
The first contribution that corresponds to diagram $(a)$ has been found in Ref. \cite{Cvetic:2006iu}. We re-present the final formula of Ref. \cite{Cvetic:2006iu}
in the following form   
\begin{eqnarray}
V^{(a)} \equiv \left[\frac{-8}{[12]^2[31]^2}  +  \frac{8}{[12][23][31]^2}  +  \frac{16}{[12][23]^2[31]}  +  \frac{8}{[12]^2[23][31]}  \right] \no \\
+ \left[\frac{2}{[12]^2[23]} + \frac{2}{[23][31]^2}  + \frac{-12}{[12][31]^2}   + \frac{-12}{[12]^2[31]}   +  \frac{-6}{[12][23]^2} 
+  \frac{-6}{[23]^2[31]}  \right.\no\\
\left. + \frac{2[12]}{[23]^2[31]^2}  + \frac{2[31]}{[12]^2[23]^2}  + \frac{8[23]}{[12]^2[31]^2}\right] J(1,1,1) \no\\
+ \left[\frac{-6}{[12]^2[23]^2} +  \frac{-4}{[12]^2[31]^2} +   \frac{2}{[23]^2[31]^2} +  \frac{-2}{[12][23][31]^2} \right.\no \\
\left. +  \frac{8}{[12][23]^2[31]} + \frac{10}{[12]^2[23][31]} \right]\ln[12] \no\\
+ \left[\frac{4}{[12]^2[23]^2} + \frac{8}{[12]^2[31]^2} + \frac{4}{[23]^2[31]^2} + \frac{-8}{[12][23][31]^2}  \right. \no\\
\left. + \frac{-16}{[12][23]^2[31]}  + \frac{-8}{[12]^2[23][31]}  \right]\ln[23] \no\\
+ \left[\frac{2}{[12]^2[23]^2} + \frac{-4}{[12]^2[31]^2} + \frac{-6}{[23]^2[31]^2} + \frac{10}{[12][23][31]^2} \right.\no \\
+  \left. \frac{8}{[12][23]^2[31]}  + \frac{-2}{[12]^2[23][31]}  \right]\ln[31] \label{aaa}
\end{eqnarray}
The total contribution  of diagram $(a)$ to the $Lcc$ correlator is this integral multiplied by a weight factor coming from combinatorics of Feynman rules, 
trivial group algebra, and Grassmanian nature of the ghost fields. The contributions  of  all the diagrams $(a), (b), (c)$ to the effective action have 
common factor $i g^4 N^2/\pi^8$ in front of them. We concentrate on the relative weights caused by degrees of $1/2$ and signs. This weight factor for diagram $(a)$ is $1/2^{16}.$

The final result for diagram $(b)$ is Eq. (\ref{b-final}). We re-present it in the following form  
\begin{eqnarray}
\frac{1}{8}\left[\frac{-3}{[12]^{2}[23]^{2}}  +  \frac{3}{[12]^{2}[31]^{2}}  + \frac{-3}{[23]^{2}[31]^{2}} + \frac{-2}{[12][23][31]^{2}} + \frac{8}{[12][23]^{2}[31]} \right] \no\\
+ \frac{1}{8}\left[\frac{-2}{[12][31]^2} +  \frac{2}{[12][23]^2} + \frac{-6}{[23]^2[31]} + \frac{4[12]}{[23]^{2}[31]^2} \right] J(1,1,1) \no\\
+\frac{1}{8} \left[\frac{-6}{[23]^2[31]^2} +  \frac{-4}{[12][23][31]^2} +  \frac{4}{[12][23]^2[31]} \right]\ln[12] \no\\
+ \frac{1}{8}\left[\frac{3}{[12]^2[23]^2} + \frac{6}{[12]^2[31]^2} + \frac{5}{[23]^2[31]^2} + \frac{-3}{[12][23][31]^2}  \right.\no\\
\left. + \frac{-8}{[12][23]^2[31]} + \frac{3}{[12]^2[23][31]}  \right]\ln[23] \no\\
+ \frac{1}{8}\left[\frac{-3}{[12]^2[23]^2} + \frac{-6}{[12]^2[31]^2} + \frac{1}{[23]^2[31]^2} + \frac{7}{[12][23][31]^2} \right.\no\\
\left. +  \frac{4}{[12][23]^2[31]}   + \frac{-3}{[12]^2[23][31]}  \right]\ln[31] \label{bbb}
\end{eqnarray}
This integral should be summed with another integral obtained from Eq. (\ref{bbb}) by exchanging $x_2$ and $x_3.$ Thus, the total contribution of diagram $(b)$  to the 
$Lcc$ correlator at two loop level is 
\begin{eqnarray}
V^{(b)} \equiv \frac{1}{8}\left[\frac{-6}{[12]^{2}[23]^{2}}  +  \frac{6}{[12]^{2}[31]^{2}}  + \frac{-6}{[23]^{2}[31]^{2}} + \frac{-2}{[12][23][31]^{2}}  \right.\no\\
\left. + \frac{16}{[12][23]^{2}[31]}  + \frac{-2}{[12]^2[23][31]}\right] + \frac{1}{8}\left[\frac{-2}{[12][31]^2} + \frac{-2}{[12]^2[31]}  + \frac{-4}{[12][23]^2}  
+  \frac{-4}{[23]^2[31]} \right.\no\\
\left. + \frac{4[12]}{[23]^{2}[31]^2}  + \frac{4[31]}{[23]^{2}[31]^2} \right] J(1,1,1) \no\\
+ \frac{1}{8}\left[\frac{1}{[12]^2[23]^2} + \frac{-6}{[12]^2[31]^2} + \frac{-9}{[23]^2[31]^2} + \frac{-7}{[12][23][31]^2} \right.\no\\
+  \left. \frac{8}{[12][23]^2[31]}  + \frac{7}{[12]^2[23][31]}  \right]\ln[12] \no\\
+ \frac{1}{8}\left[\frac{8}{[12]^2[23]^2} + \frac{12}{[12]^2[31]^2} + \frac{8}{[23]^2[31]^2} + \frac{-16}{[12][23]^2[31]}  \right]\ln[23] \no\\
+ \frac{1}{8}\left[\frac{-9}{[12]^2[23]^2} + \frac{-6}{[12]^2[31]^2} + \frac{1}{[23]^2[31]^2} + \frac{7}{[12][23][31]^2} \right.\no\\
+ \left.  \frac{8}{[12][23]^2[31]}  + \frac{-7}{[12]^2[23][31]}  \right]\ln[31] \label{bbb2}
\end{eqnarray}
The weight of diagram $(b)$ in the full result is $-1/2^{13}.$

The result for diagram $(c)$ is obtained in a different way than for diagrams $(a)$ and $(b)$ and the details of calculation of the diagram $(c)$ can be found 
in Ref. \cite{Cvetic:2007ds}. It takes the form 
\begin{eqnarray}
\left[\frac{13/2}{[12]^2[23]^2} + \frac{-33/2}{[12]^2[31]^2}  + \frac{15/2}{[23]^2[31]^2}  +  \frac{11}{[12][23][31]^2} +  \frac{-14}{[12][23]^2[31]} 
+  \frac{10}{[12]^2[23][31]}  \right]  \no\\
+ \left[\frac{-6}{[12][31]^2}  + \frac{-2}{[12]^2[31]} + \frac{6}{[12][23]^2} + \frac{-5}{[23]^2[31]} + \frac{[12]}{[23]^2[31]^2}  \right. \no\\
\left. + \frac{-2[31]}{[12]^2[23]^2} + \frac{4[23]}{[12]^2[31]^2}  \right]J(1,1,1) \no\\
+ \left[\frac{-2}{[12]^2[23]^2} +  \frac{6}{[12]^2[31]^2} +   \frac{4}{[23]^2[31]^2} +  \frac{1}{[12][23][31]^2} \right.\no\\
\left. +  \frac{-1}{[12][23]^2[31]} + \frac{-4}{[12]^2[23][31]} \right]\ln[12] \no\\
+ \left[\frac{-1/2}{[12]^2[23]^2} + \frac{-5}{[12]^2[31]^2} + \frac{-3/2}{[23]^2[31]^2} + \frac{-7/2}{[12][23][31]^2}  \right.\no\\
\left. + \frac{2}{[12][23]^2[31]} + \frac{7/2}{[12]^2[23][31]}  \right]\ln[23] \no\\
+ \left[\frac{5/2}{[12]^2[23]^2} + \frac{-1}{[12]^2[31]^2} + \frac{-5/2}{[23]^2[31]^2} + \frac{5/2}{[12][23][31]^2} \right.\no\\
\left. +  \frac{-1}{[12][23]^2[31]}  + \frac{1/2}{[12]^2[23][31]}  \right]\ln[31] \label{ccc} 
\end{eqnarray}
This integral should be summed with another integral obtained from Eq. (\ref{ccc}) by exchanging $x_2$ and $x_3.$  
Thus, the total contribution of diagram $(c)$  is 
\begin{eqnarray}
V^{(c)} \equiv \left[\frac{14}{[12]^2[23]^2} + \frac{-33}{[12]^2[31]^2}  + \frac{14}{[23]^2[31]^2}  +  \frac{21}{[12][23][31]^2} \right.\no\\
\left. +  \frac{-28}{[12][23]^2[31]} +  \frac{21}{[12]^2[23][31]}  \right] 
+ \left[\frac{-8}{[12][31]^2} + \frac{-8}{[12]^2[31]} + \frac{1}{[12][23]^2} + \frac{1}{[23]^2[31]}  \right.\no\\
\left. +  \frac{-[12]}{[23]^2[31]^2}  + \frac{-[31]}{[12]^2[23]^2}  + \frac{8[23]}{[12]^2[31]^2}  \right]J(1,1,1) \no\\
+ \left[\frac{-9/2}{[12]^2[23]^2} + \frac{5}{[12]^2[31]^2} + \frac{13/2}{[23]^2[31]^2} + \frac{3/2}{[12][23][31]^2} \right.\no\\
\left. + \frac{-2}{[12][23]^2[31]}  + \frac{-3/2}{[12]^2[23][31]} \right]\ln[12] \no\\
+ \left[\frac{-2}{[12]^2[23]^2} + \frac{-10}{[12]^2[31]^2} + \frac{-2}{[23]^2[31]^2} +  \frac{4}{[12][23]^2[31]}  \right]\ln[23] \no\\
+ \left[\frac{13/2}{[12]^2[23]^2} + \frac{5}{[12]^2[31]^2} + \frac{-9/2}{[23]^2[31]^2} + \frac{-3/2}{[12][23][31]^2} \right.\no\\
+ \left.  \frac{-2}{[12][23]^2[31]}   + \frac{3/2}{[12]^2[23][31]}  \right]\ln[31] \label{ccc2} 
\end{eqnarray}
The weight of diagram $(c)$ in the full result is $1/2^{15}.$ 

Thus, the contribution of first three diagrams $(a),(b)$ and $(c)$ to the two-loop effective action can be presented in the following form
\begin{eqnarray*}
\int~d^4x_1d^4x_2d^4x_3 \frac{i g^4 N^2}{2^{16}\pi^8} f^{def} L^d(x_1)c^e(x_2)c^f(x_3) ~ V^{(a+b+c)}(x_1,x_2,x_3),
\end{eqnarray*}
where
\begin{eqnarray*}
V^{(a+b+c)}(x_1,x_2,x_3) \equiv  V^{(a)} - 8 V^{(b)} + 2 V^{(c)} = \\
\\
\left[\frac{34}{[12]^2[23]^2} + \frac{-80}{[12]^2[31]^2}  + \frac{34}{[23]^2[31]^2}  + \frac{52}{[12][23][31]^2} + \frac{-56}{[12][23]^2[31]} 
+  \frac{52}{[12]^2[23][31]}  \right] \no\\
+ \left[\frac{2}{[12]^2[23]} + \frac{2}{[23][31]^2} + \frac{-26}{[12][31]^2} + \frac{-26}{[12]^2[31]} \right.\\
\left. + \frac{-4[12]}{[23]^2[31]^2}  + \frac{-4[31]}{[12]^2[23]^2}  + \frac{24[23]}{[12]^2[31]^2}  \right]J(1,1,1) \no\\
+ \left[\frac{-16}{[12]^2[23]^2} + \frac{12}{[12]^2[31]^2} + \frac{24}{[23]^2[31]^2} + \frac{8}{[12][23][31]^2} 
+ \frac{-4}{[12][23]^2[31]} \right]\ln[12] \no\\
+ \left[ \frac{-8}{[12]^2[23]^2} + \frac{-24}{[12]^2[31]^2} + \frac{-8}{[23]^2[31]^2} \right.\\
\left. + \frac{-8}{[12][23][31]^2}  + \frac{8}{[12][23]^2[31]}  + \frac{-8}{[12]^2[23][31]}\right]\ln[23] \no\\
+ \left[\frac{24}{[12]^2[23]^2} + \frac{12}{[12]^2[31]^2} + \frac{-16}{[23]^2[31]^2} + \frac{-4}{[12][23]^2[31]} 
+ \frac{8}{[12]^2[23][31]}  \right]\ln[31] 
\end{eqnarray*}

Diagrams $(d)$ and $(e)$  are easy to calculate since they contain propagator-like insertions only. They can produce logarithms only (due to the shift of the indices in the propagators)
and cannot produce  Davydychev integral $J(1,1,1).$  These two diagrams are divergent separately in the ultraviolet region, however in the maximally supersymmetric Yang-Mills theory 
their poles cancel each other. Thus, the contribution of diagrams $(d)$ and $(e)$ to the two-loop effective action can be presented 
in the following form
\begin{eqnarray*}
\int~d^4x_1d^4x_2d^4x_3 \frac{i g^4 N^2}{2^{16}\pi^8} f^{abc} L^a(x_1)c^b(x_2)c^c(x_3) ~ V^{(d+e)}(x_1,x_2,x_3),
\end{eqnarray*}
where
\begin{eqnarray}
V^{(d+e)} (x_1,x_2,x_3) \equiv \no\\
\left[\frac{-44/3}{[12]^2[23]^2} + \frac{304/3}{[12]^2[31]^2}  + \frac{-44/3}{[23]^2[31]^2}  + \frac{-260/3}{[12][23][31]^2} + \frac{88/3}{[12][23]^2[31]} 
+  \frac{-260/3}{[12]^2[23][31]}  \right] \no\\
\left[\frac{12}{[12]^2[23]^2} + \frac{-24}{[12]^2[31]^2} + \frac{12}{[23]^2[31]^2} + \frac{12}{[12][23][31]^2} \right.\no\\
\left. + \frac{-24}{[12][23]^2[31]}  + \frac{12}{[12]^2[23][31]} \right]\ln[12] \no\\
+\left[\frac{-24}{[12]^2[23]^2} + \frac{48}{[12]^2[31]^2} + \frac{-24}{[23]^2[31]^2} + \frac{-24}{[12][23][31]^2} \right.\no\\
\left. + \frac{48}{[12][23]^2[31]}  + \frac{-24}{[12]^2[23][31]} \right]\ln[23] \no\\
+ \left[\frac{12}{[12]^2[23]^2} + \frac{-24}{[12]^2[31]^2} + \frac{12}{[23]^2[31]^2} + \frac{12}{[12][23][31]^2} \right.\no\\
\left. + \frac{-24}{[12][23]^2[31]}  + \frac{12}{[12]^2[23][31]} \right]\ln[31] \label{d+e}
\end{eqnarray}

Thus, the full result for the two-loop $Lcc$ correlator in the maximally supersymmetric Yang-Mills theory in four spacetime dimensions 
is a combination of $V^{(a)},$ $V^{(b)},$  $V^{(c)}$  and $V^{(d+e)}$   with the correpsonding weights and has the form   
\begin{eqnarray*}
\int~d^4x_1d^4x_2d^4x_3 \frac{i g^4 N^2}{2^{15}\pi^8} f^{abc} L^a(x_1)c^b(x_2)c^c(x_3) ~ V^{(2)} (x_1,x_2,x_3),
\end{eqnarray*}
where
\begin{eqnarray*}
V^{(2)}(x_1,x_2,x_3)  =   V^{(a+b+c)}(x_1,x_2,x_3) + V^{(d+e)}(x_1,x_2,x_3) = \\
\left[\frac{29/3}{[12]^2[23]^2} + \frac{32/3}{[12]^2[31]^2}  + \frac{29/3}{[23]^2[31]^2}  + \frac{-52/3}{[12][23][31]^2} + \frac{-40/3}{[12][23]^2[31]} 
+  \frac{-52/3}{[12]^2[23][31]}  \right] \no\\
+ \left[\frac{1}{[12]^2[23]} + \frac{1}{[23][31]^2} + \frac{-13}{[12][31]^2} + \frac{-13}{[12]^2[31]}  + \frac{-2[12]}{[23]^2[31]^2} \right.\\ 
\left. +  \frac{-2[31]}{[12]^2[23]^2}  + \frac{12[23]}{[12]^2[31]^2}  \right]J(1,1,1) \no\\
+ \left[\frac{-2}{[12]^2[23]^2} + \frac{-6}{[12]^2[31]^2} + \frac{18}{[23]^2[31]^2} + \frac{10}{[12][23][31]^2} \right.\\
\left.   + \frac{-14}{[12][23]^2[31]} + \frac{6}{[12]^2[23][31]}  \right]\ln[12] \no\\
+ \left[ \frac{-16}{[12]^2[23]^2} + \frac{12}{[12]^2[31]^2} + \frac{-16}{[23]^2[31]^2} + \frac{-16}{[12][23][31]^2} \right.\\  
\left. + \frac{28}{[12][23]^2[31]}  + \frac{-16}{[12]^2[23][31]}\right]\ln[23] \no\\
+ \left[\frac{18}{[12]^2[23]^2} + \frac{-6}{[12]^2[31]^2} + \frac{-2}{[23]^2[31]^2} + \frac{6}{[12][23][31]^2} \right.\\  
\left. + \frac{-14}{[12][23]^2[31]} + \frac{10}{[12]^2[23][31]}  \right]\ln[31] 
\end{eqnarray*}

For the reference, the one-loop contribution to the effective action is 

\begin{eqnarray*}
\int~d^4x_1d^4x_2d^4x_3 \frac{i g^2 N}{2^{8}\pi^6} f^{abc} L^a(x_1)c^b(x_2)c^c(x_3) ~ V^{(1)} (x_1,x_2,x_3),
\end{eqnarray*}
where
\begin{eqnarray*}
V^{(1)}(x_1,x_2,x_3)  =  \\
\left[\frac{-1}{[12]^2[23]^2} + \frac{2}{[12]^2[31]^2}  + \frac{-1}{[23]^2[31]^2}  + \frac{-1}{[12][23][31]^2} + \frac{2}{[12][23]^2[31]} +  \frac{-1}{[12]^2[23][31]}  \right] 
\end{eqnarray*}

\section{Conclusions}
\label{sec:concl}

In this paper we presented the calculation of the second diagram [diagram  $(b)$] 
of the five planar two-loop diagrams to the $Lcc$ vertex depicted in Fig.~\ref{Lccfig}
for a general Yang-Mills theory. The Lorentz structure of the corresponding integrand is 
complicated in comparison with the integrand of the previously calculated diagram $(a)$ 
\cite{Cvetic:2006iu}. 
In order to apply the same methods as in the diagram $(a)$, we first performed in the
integrands the Lorentz algebra for certain subproducts in the numerators, then performed
one of the two $D$-dimensional integrations, then finished the Lorentz algebra, and finally
performed the second $D$-dimensional integration.
It is a long procedure that requires certain computer resources. In principle, it is 
possible to use another trick to reduce the calculation \cite{Cvetic:2007ds}, a trick which we applied 
in the calculation of diagram $(c)$ \cite{Cvetic:2007ds}. 
Neither the result for the diagram $(b)$ obtained here, nor the result for the diagram $(a)$
obtained in our previous work \cite{Cvetic:2006iu}
depend on any scale, infrared or ultraviolet,
in complete correspondence with the   naive index counting arguments, taking into account that 
none of three diagrams $(a,b,c)$ contains any divergent subgraphs due to transversality of the gauge propagator 
in the Landau gauge. From the point of view of R-operation theory the scale independence 
of this vertex in the Landau gauge in all orders of perturbation theory has been explained in Refs. \cite{Cvetic:2004kx,Kondrashuk:2004pu,Cvetic:2006kk}. 
By using the ST identity, other (leading-$N$) two-loop vertices can be derived from the
(leading-$N$) two-loop $Lcc$ vertex once the latter is known. For example,
it will be possible to derive the four-point off-shell correlator, and consequently
reproduce the known result for the four-point gluon amplitude 
\cite{Bern:2006ew,Bern:2005iz} and the anomalous dimensions of twist-two operators 
\cite{Kotikov:2000pm,Kotikov:2002ab,Kotikov:2003fb,Kotikov:2004er,Kotikov:2006ts}.

It is possible to look at these results from different points of view. 
On the one hand, singular parts of the diagrams $(d)$ and $(e)$ are proportional to the one-loop result 
for the $Lcc$ correlator written in Ref.~\cite{Cvetic:2006iu}.  
Moreover, the transversality of the gauge propagator must be conserved 
by the radiative correction, since it is a well-known fact that the gauge fixing term
does not obtain any quantum correction \cite{Faddeev:1980be}. 
This means that the sum of singular parts of diagrams $(d)$ and $(e)$ is proportional to the one-loop 
contribution, with a coefficient that is singular in a general nonsupersymmetric massless gauge theory, 
but in this particular theory it is a finite factor since the poles in $\e$ cancel 
in the sum of the diagrams $(d)$ and $(e)$ due to the ${\cal N} = 4$ supersymmetry. 
This factor is logarithm of a ratio of the spacetime intervals. This factor appears due to shift of indices by multiplies of $\e$ in 
the propagator of the fields. This shift is typical for massless theories in MS scheme.  
In the nonsupersymmetric case, 
that singular number can be absorbed into 
the gauge coupling to organize the bare coupling. Then, the bare coupling together with the logarithm of ratio 
of the distance to the scale, leads to the running (effective) coupling. It means that in $D$ dimensions the massless nonsupersymmetric 
gauge theory is a conformal gauge theory in terms 
of the running effective coupling (formed from the bare coupling) and dressed mean fields.
In other words, we think that, in terms of the running couplings and 
dressed mean fields, the arguments of Ref.~\cite{Cvetic:2004kx} can be applied without 
modifications to the massless QCD. 
The role of the renormalization group (RG) scale could be the coordinate of the moduli space 
of the theory.

On the other hand, QCD is asymptotically free theory and for short distances we can consider 
it as a theory with zero beta-function (running of the effective charge is almost absent) and 
the conformal structure could be restored at short distances
by the method of conformal theory in terms of dressed mean fields.
Further, it would be interesting to investigate the relation of off-shell
correlators of dressed mean fields and correlators of instantons in multi-dimensional theories. 
To introduce masses in the theory, we need to use softly broken 
supersymmetry, in which the couplings are spacetime-independent background superfields. 
The relation between the RG functions of softly broken and 
rigid theories was found out in Refs.~\cite{Yamada:1994id,Jack:1997pa,Avdeev:1997vx}. 
The relation between the correlators of softly broken and rigid theories
can be found by a trick of general change of variables in superspace \cite{Kondrashuk:1999de}.

To restore the conformal structure of all the effective action, we need to solve the ST identity. 
Algorithm for solving the ST identity could be applied to various theories, such as   
the ${\cal N}=8$ supergravity \cite{Bjerrum-Bohr:2006yw,Kang:2004cs,Green:2006gt,Bern:2006kd}, 
Chern-Simons theory near the RG fixed points \cite{Avdeev:1992jt}, 
massless gauge theory near fixed points in the coupling space, 
topological field theories in higher dimensions, finite ${\cal N}=1$ supersymmetric theories 
\cite{Jones:1986vp,Ermushev:1986cu,Kazakov:1991th,Kazakov:1995cy,Kondrashuk:1997uf}.  
We further note that, in addition to the five planar two-loop diagrams of 
Fig.~\ref{Lccfig}, there is one nonplanar diagram (the nonplanar variant of
the diagram $(a)$)
which is suppressed in the planar limit of the large 't Hooft coupling 
by the factor $1/N.$

\subsection*{Acknowledgments}

The work of I.K. was supported by Ministry of Education (Chile) under grant Mecesup FSM9901 
and by DGIP UTFSM, by  Fondecyt (Chile) grant \#1040368, 
and by Departamento de Ciencias B\'asicas de la Universidad del B\'\i o-B\'\i o, Chill\'an 
(Chile). The work of G.C. was supported in part by Fondecyt (Chile) grant \#1050512. 
We are grateful to Anatoly Kotikov for checking a part of formulas in the manuscript 
and to Ivan Schmidt for careful reading of the manuscript.

\subsection*{Note added}

After publication of the first version of this paper the ``dual conformal  symmetry'' of the momentum space 
(conformal symmetry of four-point correlator of gluon in the momentum space) 
found in Ref. \cite{Drummond:2006rz}  has been exploited further in Ref. \cite{Nguyen:2007ya}
up to four loop level to detect all off-shell correlators that will contribute to 
four-point amplitude. That symmetry appears also on the strong coupling side as it has been 
found by Alday and Maldacena in Ref. \cite{Alday:2007hr} via AdS/CFT correspondence. As we have already 
mentioned in  Introduction, Fourier transform of the  UD integrals of the momentum space can be expressed  
in terms of UD integrals in the position space. This observation could be a bridge between the dual conformal symmetry 
of the momentum space and conformal symmetry of the effective action of dressed mean fields 
in the position space.

\begin{appendix}
\section[]{}
\setcounter{equation}{0}
 \label{App:A}
In this Appendix we present the result for the second and the third integrals 
(\ref{Ib2def}) and (\ref{Ib3def}) in terms of $D$-dimensional integrals $J$. 
The result for the second integral is
\begin{eqnarray}
I_{b,2} = 
\frac{A(1,1-2\e,1)}{8\e(1-2\e)^2}
\left\{\frac{(8-20\e)(1-\e^2)}{[31][12]^{-\e}} J(2+\e,2-3\e,1) \right.\no\\ 
+ \left. \frac{(1-\e^2)(4-6\e)}{[31][12]^{-1-\e}} J(2+\e,3-3\e,1) \right. \no\\
+ \left.    \left[-\frac{2\e^2(1-\e)}{[31][12]^{1-\e}} 
+ \frac{2(2-\e-4\e^2)(1-\e)}{[31]^2[12]^{-\e}}\right] J(\e,3-3\e,1) \right.\no\\  
+ \left. \frac{2(1-\e^2)(2-3\e)}{[31][12]^{1-\e}} J(2+\e,1-3\e,1) \right. \no\\
- \left. \frac{2(2-\e-4\e^2)(1-\e)}{[31]^2[12]^{1-\e}} J(1+\e,2-3\e,0)   
+ \frac{2(1-\e^2)(2-3\e)}{[31]^2[12]^{1-\e}} J(2+\e,1-3\e,0) \right.\no\\
- \left. \frac{2\e^2(1-\e)}{[31]^2[12]^{1-\e}} J(\e,3-3\e,0) \right. \no\\
+ \left. \left[-\frac{7-17\e-8\e^2+20\e^3}{[31]^2[12]^{-\e}} 
- \frac{2(2-\e-4\e^2)(1-\e)}{[31][12]^{1-\e}}\right] J(1+\e,2-3\e,1) \right. \no\\
+ \left. \left[-\frac{2(1-\e^2)(2-3\e)}{[31]^2[12]^{-1-\e}} 
- \frac{2(2-\e-4\e^2)(1-\e)}{[31][12]^{-\e}}\right] J(1+\e,3-3\e,1) \right. \no\\
+ \left. \frac{4-5\e-6\e^2+8\e^3}{[31]^2[12]^{1-\e}} J(\e,2-3\e,1)  \right.\no\\
+ \left. \frac{2\e^2(1-\e)}{[31]^2[12]^{1-\e}} J(\e-1,3-3\e,1)  
- \frac{(1-2\e^2)(2-3\e)}{[31]^2[12]^{1-\e}} J(1+\e,1-3\e,1)\right. \no\\
+ \left. \le \frac{\e(2-\e)[21]}{[31][23]^{1-\e}} 
- \frac{(2-\e)(2-3\e)}{[31][23]^{-\e}}    \ri   J(2,2-3\e,1+\e)  \right.\no\\
+ \left.\le \frac{(\e^2-2\e)[21]}{[31]^2[23]^{1-\e}} 
+ \frac{4-8\e+3\e^2}{[31]^2[23]^{-\e}}  
+ \frac{\e^2-\e}{[31][23]^{1-\e}}\ri J(1,2-3\e,1+\e) \right.\no\\
+ \left. \le \frac{4-10\e + 6\e^2}{[31]^2[23]^{-\e}}  
-  \frac{\e(1-2\e)}{[31][23]^{1-\e}}   
+   \frac{(3\e-4\e^2)[21]}{[31]^2[23]^{1-\e}}   \ri J(1,3-3\e,\e) \right.\no\\
- \left.  \frac{(2-\e)(2-3\e)[21]}{[31][23]^{-\e}}J(2,3-3\e,1+\e) \right.\no\\ 
+ \left. \le \frac{(2-\e)(2-3\e)[21]}{[31]^2[23]^{-\e}} 
+  \frac{(2-\e)(2-3\e)}{[31][23]^{-\e}}  \ri   J(1,3-3\e,1+\e) \right.\no\\  
+ \left. \left[\frac{\e(1+\e)[21]}{[31]^2[23]^{1-\e}} 
+ \frac{\e(1-\e)[21]}{[31][23]^{2-\e}}  
- \frac{[12]^2\e(1-\e)}{[31]^2[23]^{2-\e}}   \right.\right.\no\\
- \left.\left. \frac{2(1-\e)(2-3\e)}{[31]^2[23]^{-\e}}  
+  \frac{\e(1-2\e)}{[31][23]^{1-\e}}   \right] J(2,2-3\e,\e) \right.   \no\\
+ \left. \left[-\frac{2\e(1-\e)[21]}{[31]^2[23]^{1-\e}} 
- \frac{\e(1-\e)[21]}{[31][23]^{2-\e}}  
+ \frac{[12]^2 \e(1-\e)}{[31]^2[23]^{2-\e}} \right] J(2,3-3\e,\e-1)   \right.  \no\\ 
+ \left. \left[-\frac{2(2-3\e)(1-\e)[21]}{[31]^2[23]^{-\e}} 
+ \frac{\e(1-2\e)[21]}{[31][23]^{1-\e}}  
- \frac{[12]^2\e(2-3\e)}{[31]^2[23]^{1-\e}} \right] J(2,3-3\e,\e) \right. \no \\
+ \left. \frac{1-3\e}{[21]^{-\e}[31]} J(1+\e,2-3\e,2) 
+ \le -\frac{1-3\e}{[31]^2[21]^{-\e}} 
+ \frac{\e}{[31][21]^{1-\e}} \ri J(\e,2-3\e,2)  \right.\no\\ 
- \left. \frac{\e}{[31]^2[21]^{1-\e}} J(\e-1,2-3\e,2) \right.\no\\
+ \left.  \frac{2-3\e}{[31][21]^{1-\e}} J(1+\e,1-3\e,2) 
- \frac{2-3\e}{[31]^2[21]^{1-\e}}J(\e,1-3\e,2)  \right.\no\\
+ \left. \frac{\e(1-\e)}{[31][23]^{1-\e}} J(2,1-3\e,1+\e) \right.\no\\
+ \left. \le - \frac{(1-2\e)(1-\e)}{[31]^2[23]^{1-\e}}  
- \frac{(1-\e)^2}{[31][23]^{2-\e}} 
+ \frac{(1-\e)^2[12]}{[31]^2[23]^{2-\e}} \ri J(2,1-3\e,\e) \right.\no \\
- \left.  \frac{\e(1-\e)}{[31]^2[23]^{1-\e}}J(1,1-3\e,1+\e)  \right. \no\\
+  \left.  \le - \frac{2\e(1-\e)}{[31]^2[23]^{1-\e}} 
-  \frac{\e(1-\e)}{[31][23]^{2-\e}}  +    \frac{\e(1-\e)[12]}{[31]^2[23]^{2-\e}}    
\ri  J(2,2-3\e,\e-1)  \right. \no\\
+ \left. \le \frac{(1-\e)^2}{[31]^2[23]^{1-\e}} 
+  \frac{(1-\e)^2}{[31][23]^{2-\e}}  - \frac{(1-\e)^2[21]}{[31]^2[23]^{2-\e}}    
\ri   J(1,2-3\e,\e)  \right.   \no\\  
+ \left.  \frac{\e(1-\e)}{[31]^2[23]^{1-\e}}J(0,2-3\e,1+\e)   \right.\no\\
+ \left.  \le \frac{2\e(1-\e)}{[31]^2[23]^{1-\e}} + \frac{\e(1-\e)}{[31][23]^{2-\e}} 
- \frac{\e(1-\e)[12]}{[31]^2[23]^{2-\e}}  \ri J(1,3-3\e,\e-1)\right. \no\\
- \left.  \frac{\e(1-\e)}{[31]^2[23]^{1-\e}} J(0,3-3\e,\e)  \right. \no\\ 
+ \left. \le \frac{1-\e}{[31][23]^{2-\e}} + \frac{1-\e}{[31]^{2}[23]^{1-\e}}  
- \frac{(1-\e)[12]}{[31]^{2}[23]^{2-\e}} \ri J(2,1-2\e,0) \right. \no\\
- \left.  \le \frac{1-\e}{[31][23]^{2-\e}} + \frac{1-\e}{[31]^{2}[23]^{1-\e}}  
- \frac{(1-\e)[12]}{[31]^{2}[23]^{2-\e}} \ri J(1,2-2\e,0)\right\}  \ . \label{second}
\end{eqnarray}
The third double integral (\ref{Ib3def}) is simpler, 
the integration over $y$ can be taken immediately,
\begin{eqnarray}
I_{b,3} = 
\frac{A(1,1-2\e,1)}{4\e(1-2\e)}
\left\{-\frac{1-2\e^2}{[12]^{1-\e}[31]^2}J(1+\e,1-3\e,1) \right. \no\\
 \left. \le-\frac{1+\e-4\e^2}{[12]^{-\e}[31]^2}  
+ \frac{(2\e-2\e^2)[23]}{[12]^{1-\e}[31]^2} 
+ \frac{1-2\e-2\e^2}{[12]^{1-\e}[31]}\ri J(1+\e,2-3\e,1) \right. \no \\
+  \left. \frac{1+2\e-4\e^2}{[12]^{1-\e}[31]^2} J(\e,2-3\e,1)  
-  \frac{2+2\e-4\e^2}{[12]^{1-\e}[31]^2} J(1+\e,2-3\e,0)  \right. \no\\
- \left. \frac{2\e-2\e^2}{[12]^{-\e}[31]^2} J(1+\e,3-3\e,0) 
+ \frac{2\e-2\e^2}{[12]^{1-\e}[31]^2} J(\e,3-3\e,0)  \right. \no\\
 + \left. \le\frac{(2\e-2\e^2)[23]}{[12]^{-\e}[31]^2}  
-  \frac{\e}{[12]^{-\e}[31]} 
- \frac{\e - 2\e^2}{[12]^{-\e-1}[31]^2} \ri J(1+\e,3-3\e,1) \right. \no\\
+ \left. \frac{2+\e}{[12]^{1-\e}[31]}J(1+\e,1-3\e,2)   \right. \no\\
+  \left. \le -\frac{2+\e}{[12]^{1-\e}[31]}  
- \frac{\e[23]}{[12]^{1-\e}[31]^2}  
- \frac{1}{[12]^{-\e}[31]^2}\ri J(\e,2-3\e,2) \right. \no\\
+   \left.   \le  - \frac{2\e(1-\e)[23]}{[12]^{1-\e}[31]^2}  
+  \frac{\e}{[12]^{1-\e}[31]} 
+ \frac{3\e-4\e^2}{[12]^{-\e}[31]^2} \ri J(\e,3-3\e,1) \right. \no\\
+  \left.  \frac{\e[23]}{[12]^{-\e}[31]}J(1+\e,3-3\e,2)  
-  \le\frac{\e[23]}{[12]^{1-\e}[31]} 
+ \frac{\e[23]}{[12]^{-\e}[31]^2} \ri J(\e,3-3\e,2) \right.\no\\
- \left. \frac{1}{[12]^{1-\e}[31]^2} J(\e,1-3\e,2)   
+ \frac{1}{[12]^{1-\e}[31]^2} J(\e-1,2-3\e,2)  \right. \no\\
 \left.  -  \frac{2\e-2\e^2}{[12]^{1-\e}[31]^2} J(\e-1,3-3\e,1)  
+ \frac{\e[23]}{[12]^{1-\e}[31]^2} J(\e-1,3-3\e,2)  \right. \no\\
+ \left. \frac{2-2\e^2}{[12]^{1-\e}[31]^2} J(2+\e,1-3\e,0)   \right. \no\\
\left. - \frac{1+\e}{[12]^{1-\e}}  J(2+\e,1-3\e,2) 
- \frac{1+\e}{[12]^{-\e}}J(2+\e,2-3\e,2)  \right.\no\\
+ \left.  \le\frac{1+\e}{[12]^{1-\e}}  + \frac{2+\e}{[12]^{-\e}[31]}  
+   \frac{\e[23]}{[12]^{1-\e}[31]}\ri J(1+\e,2-3\e,2) \right. \no\\
- \left. \frac{1-\e-2\e^2}{[12]^{1-\e}[31]} J(2+\e,1-3\e,1)  
- \frac{1-\e-2\e^2}{[12]^{-\e}[31]} J(2+\e,2-3\e,1) \right\} \ . \label{third}
\end{eqnarray}

In Eqs. (\ref{second}) and (\ref{third}), we do not write single integrals with one of the indices zero and sum of other two indices equal to $D,$ $D+1$ or $D+2.$
Such terms are proportional to delta-functions in position space multiplied by pole in $\e$ and will disappear in the final expression (\ref{b-final}).

\section[]{}
\setcounter{equation}{0}
\label{App:B}

In this Appendix we write the results for the terms contributing to the diagram $(b)$ 
with $J$'s whose sum of indices is $k=D, D+1, D+2$. These terms appear in $I_b$ which
is the sum of Eqs. (\ref{idn}), (\ref{second}) and (\ref{third}). 
The integrals in $I_b^{(D)}$ are calculated by using the uniqueness method
\begin{eqnarray}
I_b^{(D)} =
\frac{A(1,1-2\e,1)}{4\e(1+2\e)(1-2\e)^2}
\left\{\le\frac{10-18\e-12\e^2}{[31]^2[12]^{-\e}} 
+ \frac{3-4\e-10\e^2}{[31][12]^{1-\e}}  + \frac{(2\e-2\e^2)[23]}{[31]^2[12]^{1-\e}}\ri 
J(1+\e,2-3\e,1) \right.\no\\
+ \left. \frac{-3+3\e+12\e^2}{[31][12]^{1-\e}} J(2+\e,1-3\e,1)  
+  \frac{2\e-4\e^2}{[31]^2[12]^{1-\e}} J(2+\e,1-3\e,0)    \right.  \no\\
+ \left. \frac{-4\e+8\e^2}{[31]^{2}[12]^{1-\e}}J(1+\e,2-3\e,0)  
+ \frac{-2\e + 2\e^2}{[31][23]^{1-\e}} J(2,1-3\e,1+\e)  \right. \no \\
+ \left. \left[\frac{-8 + 16\e - 6\e^2}{[31]^2[23]^{-\e}} 
+ \frac{-\e + 3\e^2}{[31][23]^{1-\e}}+
\frac{(-2\e + 5\e^2)[21]}{[31]^2[23]^{1-\e}}\right]J(1,2-3\e,1+\e) \right.\no  \\
+ \left. \left[\frac{(7\e-14\e^2)[21]}{[31]^2[23]^{1-\e}} 
+ \frac{(\e-3\e^2)[21]}{[31][23]^{2-\e}}  
+ \frac{[12]^2(-\e+3\e^2)}{[31]^2[23]^{2-\e}} \right.\right.\no\\ 
+ \left.\left. \frac{8-20\e+12\e^2}{[31]^2[23]^{-\e}}  
+  \frac{-2\e + 4\e^2)}{[31][23]^{1-\e}}   \right] J(2,2-3\e,\e) \right.\no\\
+ \left. \frac{1+2\e-6\e^2}{[12]^{1-\e}[31]} J(1+\e,1-3\e,2)  
+ \le \frac{-2+2\e}{[31][23]^{2-\e}} + \frac{-2+2\e}{[31]^{2}[23]^{1-\e}}  
+ \frac{[12](2-2\e)}{[31]^{2}[23]^{2-\e}} \ri J(2,1-2\e,0) \right.\no \\
+ \left. \frac{2\e-4\e^2}{[12]^{1-\e}[31]^2} J(\e,3-3\e,0)  
+ \le \frac{-2+6\e^2}{[12]^{1-\e}[31]}  
+ \frac{(-\e)[23]}{[12]^{1-\e}[31]^2}  
+ \frac{-2+5\e-2\e^2}{[12]^{-\e}[31]^2}\ri J(\e,2-3\e,2) \right. \no\\
+ \left. \le \frac{(-2\e+2\e^2)[23]}{[12]^{1-\e}[31]^2}  
+  \frac{\e-2\e^2}{[12]^{1-\e}[31]} 
+ \frac{4-11\e+2\e^2}{[12]^{-\e}[31]^2} \ri J(\e,3-3\e,1) \right. \no\\
+ \left. \le \frac{4-18\e + 26\e^2}{[31]^2[23]^{-\e}}  
+  \frac{-\e +4\e^2}{[31][23]^{1-\e}} 
+ \frac{(3\e-10\e^2)[21]}{[31]^2[23]^{1-\e}}   \ri J(1,3-3\e,\e) \right.\no\\
+ \left. \frac{(\e)[23]}{[12]^{1-\e}[31]^2} J(\e-1,3-3\e,2)  \right. \no\\
+ \left. \left[\frac{(-2\e+6\e^2)[21]}{[31]^2[23]^{1-\e}} 
+ \frac{(-\e+3\e^2)[21]}{[31][23]^{2-\e}}  
+ \frac{[12]^2 (\e-3\e^2)}{[31]^2[23]^{2-\e}} \right] J(2,3-3\e,\e-1)  \right.  \no\\ 
+ \left.  \frac{\e-3\e^2}{[31]^2[23]^{1-\e}}J(0,2-3\e,1+\e)  
+  \frac{-\e+3\e^2}{[31]^2[23]^{1-\e}}  J(0,3-3\e,\e)  \right. \no\\ 
+ \left.  \le \frac{-1+3\e-2\e^2}{[31][23]^{2-\e}} 
+ \frac{-1+3\e-2\e^2}{[31]^{2}[23]^{1-\e}}  
+ \frac{(1-3\e+2\e^2)[12]}{[31]^{2}[23]^{2-\e}} \ri J(1,2-2\e,0)\right\} \no\\
= \frac{A(1,1-2\e,1)A(1+\e,1-3\e,1)}{8\e^2(1+2\e)(1-2\e)^2}
\left\{\frac{-2\e+10\e^2}{[12]^{1-2\e}[23]^{1-2\e}[31]^{2+2\e}}  \right.\no\\
+ \left.  \frac{3+7\e+3\e^2}{[12]^{2-2\e}[23]^{1-2\e}[13]^{1+2\e}}  
+ \frac{-5+8\e-18\e^2}{[12]^{2-2\e}[23]^{-2\e}[13]^{2+2\e}} 
+  \frac{-6\e+\e^2}{[31]^2[12]^{2-2\e}} \right.\no\\
+ \left.  \frac{2-4\e-5\e^2}{[12]^{1-2\e}[23]^{2-2\e}[31]^{1+2\e}} 
+ \frac{2-4\e}{[12]^{-2\e}[23]^{2-2\e}[31]^{2+2\e}}  
+ \frac{\e - (1/2)\e^2}{[12]^{2-2\e}[23]^{2-2\e}[31]^{2\e}}  
+  \frac{\e+\e^2/2}{[31]^2[23]^{2-2\e}} \right\}\no\\
+ \frac{A^2(1,1-2\e,1)}{4\e^2(1+2\e)(1-2\e)^2}
\left\{\frac{1-\e}{[12]^{1-\e}[23]^{2-\e}[31]} 
+ \frac{1-\e}{[12]^{1-\e}[23]^{1-\e}[31]^2} - 
\frac{1 - \e}{[12]^{-\e}[23]^{2-\e}[31]^2} \right\}. \label{D}
\end{eqnarray} 
The terms in the sums $I_b^{(D+1)}$ and $I_b^{(D+2)}$ can be calculated also by using
the uniqueness method, but only after having represented the $J$-terms as derivatives
of integrals $J(\alpha_1,\alpha_2,\alpha_3)$ with $\sum \alpha_j = D$.
\begin{eqnarray}
I_b^{(D+1)}+I_b^{(D+2)} =
\frac{A(1,1-2\e,1)}{4\e(1+2\e)(1-2\e)^2}
\left\{\frac{-11+13\e+32\e^2}{[31][12]^{-\e}}J(2+\e,2-3\e,1)  \right.\no\\
+ \left. \left[\frac{8-16\e+6\e^2}{[31][23]^{-\e}} 
+ \frac{(2\e-5\e^2)[21]}{[31][23]^{1-\e}}\right] J(2,2-3\e,1+\e) \right.\no\\
+ \left.  \le\frac{(2\e-2\e^2)[23]}{[12]^{-\e}[31]^2}  
+  \frac{-4 +13\e - 6\e^2}{[12]^{-\e}[31]} 
+ \frac{-4 +13\e -6\e^2}{[12]^{-\e-1}[31]^2} \ri J(1+\e,3-3\e,1) \right. \no\\
+ \left. \frac{(\e)[23]}{[12]^{-\e}[31]}J(1+\e,3-3\e,2)  
-  \le\frac{(\e)[23]}{[12]^{1-\e}[31]} 
+ \frac{(\e)[23]}{[12]^{-\e}[31]^2} \ri J(\e,3-3\e,2) \right.\no\\
+ \left.  \frac{-1-\e+4\e^2}{[12]^{1-\e}}  J(2+\e,1-3\e,2) 
+ \frac{-1-\e+4\e^2}{[12]^{-\e}}J(2+\e,2-3\e,2)  \right.\no\\
+ \left.  \le\frac{1+\e-4\e^2}{[12]^{1-\e}}  
+ \frac{3-4\e-2\e^2}{[12]^{-\e}[31]}  
+   \frac{(\e)[23]}{[12]^{1-\e}[31]}\ri J(1+\e,2-3\e,2) \right.\no \\
+ \left. \frac{4-14\e+8\e^2}{[31][12]^{-1-\e}} J(2+\e,3-3\e,1) \right. \no\\
+ \left.  \frac{(-4+16\e-19\e^2)[21]}{[31][23]^{-\e}}J(2,3-3\e,1+\e) \right. \no\\
+ \left.\le \frac{(4-16\e+19\e^2)[21]}{[31]^2[23]^{-\e}} 
+  \frac{4-16\e+19\e^2}{[31][23]^{-\e}}  \ri   J(1,3-3\e,1+\e) \right.\no\\  
+ \left. \left[\frac{(-4+18\e-26\e^2)[21]}{[31]^2[23]^{-\e}} 
+ \frac{(\e-4\e^2)[21]}{[31][23]^{1-\e}}  
+ \frac{[12]^2(-2\e+7\e^2)}{[31]^2[23]^{1-\e}} \right] J(2,3-3\e,\e) \right\} \no\\
= \frac{A(1,1-2\e,1)A(1-3\e,1,1+\e)}{8\e^2(1+2\e)(1-2\e)^2}
\left\{\frac{2-(17/2)\e-(35/4)\e^2}{[12]^{1-2\e}[23]^{1-2\e}[13]^{2+2\e}} \right.\no\\
+ \left. \frac{-3-(3/2)\e-(7/4)\e^2}{[12]^{2-2\e}[23]^{1-2\e}[13]^{1+2\e}} 
+ \frac{5 -5\e+ (39/2)\e^2}{[12]^{2-2\e}[23]^{-2\e}[13]^{2+2\e}}  \right.\no\\
+ \left. \frac{1-4\e+(21/2)\e^2}{[12]^{1-2\e}[23]^{2-2\e}[31]^{1+2\e}} 
+ \frac{-4+(15/2)\e -(19/4)\e^2}{[12]^{-2\e}[23]^{2-2\e}[31]^{2+2\e}} 
+  \frac{(5/2)\e-(7/4)\e^2}{[12]^{2-2\e}[23]^{2-2\e}[31]^{2\e}}  \right\} \ . \label{D+1}
\end{eqnarray}

\section{}
\setcounter{equation}{0}
\label{App:C}

In this Appendix we provide formulas used in Sec.~\ref{section(D-1)}.
The following formula results from applying the corresponding IBP procedure:
\begin{eqnarray}
J(1,3-3\e,\e-1) = -\frac{1}{2-3\e}\left[J(2,2-3\e,\e-1) 
-  [12] J(2,3-3\e,\e-1)  \right. \no\\ 
+ \left. (\e-1)\le J(1,2-3\e,\e) -   [23] J(1,3-3\e,\e) \ri \right] = \no\\
- \frac{1}{2-3\e}\left[J(2,2-3\e,\e-1) 
- \frac{A(2,3-3\e,\e-1)}{[12]^{2-2\e}[23]^{-\e}[31]^{-1+2\e}} \right. \no\\ 
+ \left. (\e-1)\le J(1,2-3\e,\e) 
- \frac{A(1,3-3\e,\e)}{[12]^{2-2\e}[23]^{-\e}[31]^{-1+2\e}}\ri \right] \ . \label{aux1}
\end{eqnarray}
Other useful formulas can be taken from Ref.~\cite{Cvetic:2006iu}
(Appendix A there, obtained by IBP procedure):
\begin{eqnarray}
J(1-3\e,2,\e) 
= -(1-3\e)J(2-3\e,1,\e) - \e J(1-3\e,1,1+\e)   \no\\
- \frac{A(1,1-3\e,1+\e)}{2\e} \frac{1}{[12]^{1-2\e}[13]^{-\e}[23]^{2\e}},  \label{aux2}\\
J(2,1-3\e,\e) = -(1-3\e)J(1,2-3\e,\e) - \e J(1,1-3\e,1+\e)   \no\\
- \frac{A(1,1-3\e,1+\e)}{2\e} \frac{1}{[12]^{1-2\e}[23]^{-\e}[31]^{2\e}}, \label{aux3} \\
 J(\e,1-3\e,2) = -(1-3\e)J(\e,2-3\e,1) - \e J(1+\e,1-3\e,1)  \no \\
- \frac{A(1,1-3\e,1+\e)}{2\e} \frac{1}{[23]^{1-2\e}[12]^{-\e}[31]^{2\e}} \ . \label{aux4}
\end{eqnarray}
The following useful relation can be directly derived:
\begin{eqnarray} 
J(2,2-3\e,\e-1) \equiv \int~Dx\frac{1}{[x1]^2[x2]^{2-3\e}[x3]^{\e-1}} = 
\int~Dx\frac{[x3]}{[x1]^2[x2]^{2-3\e}[x3]^{\e}} \no\\
= \int~Dx\frac{[x1]+[13]+2(x1)(13)}{[x1]^2[x2]^{2-3\e}[x3]^{\e}} \no\\
=  J(1,2-3\e,\e) + [31]J(2,2-3\e,\e) 
+ 2(13)\int~Dx\frac{(x1)}{[x1]^2[x2]^{2-3\e}[x3]^{\e}} \no\\ 
=  J(1,2-3\e,\e) + [31]J(2,2-3\e,\e) 
+ (13)_\mu \pd_{\mu}^{(1)}J(1,2-3\e,\e) \no\\
=  J(1,2-3\e,\e) + \frac{A(2,2-3\e,\e)}{[12]^{2-2\e}[23]^{-\e}[31]^{-1+2\e}} 
+ (13)_\mu \pd_{\mu}^{(1)}J(1,2-3\e,\e) \no\\
=  J(1,2-3\e,\e) - \frac{(1-2\e)}{2\e(1-3\e)}
\frac{A(1,1-3\e,1+\e)}{[12]^{2-2\e}[23]^{-\e}[31]^{-1+2\e}} 
+ (13)_\mu \pd_{\mu}^{(1)}J(1,2-3\e,\e), \no\\
J(\e-1,2-3\e,2) = \no\\
=  J(\e,2-3\e,1) - \frac{(1-2\e)}{2\e(1-3\e)}
\frac{A(1,1-3\e,1+\e)}{[23]^{2-2\e}[12]^{-\e}[31]^{-1+2\e}} 
+ (31)_\mu \pd_{\mu}^{(3)}J(\e,2-3\e,1). \label{aux5}
\end{eqnarray} 
For $J(2 - 3 \e, 1, \e)$ we use specific formulas of Ref.~\cite{Cvetic:2006iu}
which were obtained by applying the GPT:
\begin{eqnarray} 
J(2-3\e,1,\e) = \frac{1}{[12]^{1-\e}} A(1- 3\e,1,1+\e) \frac{1}{2(1-3\e)} \times \no\\
\left[\frac{1}{\e} - \ln\frac{[13]}{[12]} 
- \e \le\frac{1}{2}\ln\frac{[12][13]}{[23]^2}\ln\frac{[13]}{[12]} 
+ ([12] + [23] - [13])J(1,1,1) \ri\right], \no\\
J(1,2-3\e,\e) = \frac{1}{[12]^{1-\e}} A(1- 3\e,1,1+\e) \frac{1}{2(1-3\e)} \times \no\\
\left[\frac{1}{\e} - \ln\frac{[23]}{[12]} 
- \e \le\frac{1}{2}\ln\frac{[12][23]}{[13]^2}\ln\frac{[23]}{[12]} 
+ ([12] + [13] - [23])J(1,1,1) \ri\right], \no\\
J(\e,2-3\e,1) = \frac{1}{[23]^{1-\e}} A(1- 3\e,1,1+\e) \frac{1}{2(1-3\e)} \times \no\\
\left[\frac{1}{\e} - \ln\frac{[12]}{[23]} 
- \e \le\frac{1}{2}\ln\frac{[23][12]}{[13]^2}
\ln\frac{[12]}{[23]} + ([23] + [13] - [12])J(1,1,1) \ri\right]. \label{aux6}
\end{eqnarray} 

\section{}
\setcounter{equation}{0}
\label{App:D}

In this appendix we apply formulas (\ref{aux6})and
obtain for the term in Eq.~(\ref{idn4}) proportional to 
$\partial_{\mu}^{(1)} J(1, 2 - 3 \e, \e)$
\begin{eqnarray}
I_b^{(\partial J,1)} \equiv
\frac{A(1,1-2\e,1)}{4\e(1+2\e)(1-2\e)^2}
\left[\frac{3\e-(5/2)\e^2}{[31]^2[23]^{1-\e}} + \frac{(3/2)\e-(5/4)\e^2}{[31][23]^{2-\e}}  
+ \frac{[12](-(3/2)\e+(5/4)\e^2)}{[31]^2[23]^{2-\e}} \right]\times\no\\
\times (13)_\mu \pd_{\mu}^{(1)}J(1,2-3\e,\e) = \no\\
- \frac{A(1,1-2\e,1)A(1- 3\e,1,1+\e)}{8\e(1+2\e)(1-2\e)^2}
\left[\frac{9/2}{[12]^{1-\e}[31]^2[23]^{1-\e}} + \frac{3/2}{[12]^{2-\e}[31][23]^{1-\e}} 
\right.\no\\
+ \left. \frac{-3}{[12]^{2-\e}[31]^2[23]^{-\e}}  
+  \frac{3/2}{[12]^{2-\e}[23]^{2-\e}}  + \frac{-3/2}{[12]^{-\e}[31]^2[23]^{2-\e}} \right]  \no\\
- \frac{1}{8}\left[\frac{21/4}{[12][23][31]^2} 
+ \frac{7/4}{[12]^{2}[23][31]} + \frac{-7/2}{[12]^2[31]^2} +  \frac{7/4}{[12]^{2}[23]^{2}}  
+ \frac{-7/4}{[31]^2[23]^{2}} \right] \no\\
- \frac{3}{16}\left[\frac{2}{[31]^2[23]} + \frac{1}{[31][23]^{2}}  
- \frac{[12]}{[31]^2[23]^{2}} \right] (13)_\mu \pd_{\mu}^{(1)} 
\frac{1}{[12]} \ln\frac{[23]}{[12]}, 
\label{thirdline} 
\end{eqnarray}
 and for the term in Eq.~(\ref{idn4}) proportional to $\partial_{\mu}^{(3)} J(\e, 2 - 3 \e, 1)$
\begin{eqnarray}
I_b^{(\partial J,2)} \equiv
\frac{A(1,1-2\e,1)}{4 \e(1+2\e)(1-2\e)^2} 
\frac{\left( 1-(5/2)\e^2 \right)}{[12]^{1-\e}[31]^2} (31)_{\mu}
\partial_{\mu}^{(3)} J(\e, 2 - 3 \e, 1) 
\no\\
=\frac{A(1,1-2\e,1)}{4\e(1+2\e)(1-2\e)^2}
\frac{\left( 1-(5/2)\e^2 \right)}{[12]^{1-\e}[31]^2} (31)_\mu \pd_{\mu}^{(3)}
\left\{
\frac{1}{[23]^{1-\e}} A(1- 3\e,1,1+\e) \frac{1}{2(1-3\e)} \times \right. \no\\
\left. \left[\frac{1}{\e} - \ln\frac{[12]}{[23]} 
- \e \le\frac{1}{2}\ln\frac{[23][12]}{[13]^2}\ln\frac{[12]}{[23]} 
+ ([23] + [13] - [12])J(1,1,1) \ri\right]  
\right\} \no\\
= -\frac{A(1,1-2\e,1)A(1- 3\e,1,1+\e)}{8\e^2(1+2\e)(1-2\e)^2}
\left[\frac{1}{[12]^{1-\e}[23]^{2-\e}[31]} \right.\no\\
+ \left. \frac{1}{[12]^{1-\e}[23]^{1-\e}[31]^2}   
- \frac{1}{[12]^{-\e}[23]^{2-\e}[31]^2}  \right]  \no\\
+ \frac{A(1,1-2\e,1)A(1- 3\e,1,1+\e)}{8\e(1+2\e)(1-2\e)^2} 
\left[\frac{1}{[12]^{1-\e}[23]^{2-\e}[31]} \right.\no\\
+ \left. \frac{1}{[12]^{1-\e}[23]^{1-\e}[31]^2} 
- \frac{1}{[12]^{-\e}[23]^{2-\e}[31]^2} \right]\ln\frac{[12]}{[23]}  \no\\
- \frac{A(1,1-2\e,1)A(1- 3\e,1,1+\e)}{8\e(1+2\e)(1-2\e)^2} 
\left[\frac{1}{[12]^{1-\e}[23]^{2-\e}[31]} \right.\no\\
+ \left. \frac{1}{[12]^{1-\e}[23]^{1-\e}[31]^2} 
-  \frac{1}{[12]^{-\e}[23]^{2-\e}[31]^2}      \right]  \no\\
- \frac{1}{8}\frac{1}{[12][31]^2} (31)_\mu \pd_{\mu}^{(3)}
\frac{1}{[23]}\le\frac{1}{2}\ln\frac{[23][12]}{[13]^2}\ln\frac{[12]}{[23]} 
+ ([23] + [13] - [12])J(1,1,1) \ri \no \\
- \frac{1}{16}\left[\frac{1}{[12][23]^{2}[31]} + \frac{1}{[12][23][31]^2} 
-   \frac{1}{[23]^{2}[31]^2}  \right] \no\\   
+ \frac{1}{4} \left[\frac{1}{[12][23]^{2}[31]} + \frac{1}{[12][23][31]^2} 
-  \frac{1}{[23]^{2}[31]^2} \right]\ln\frac{[12]}{[23]} \ .    
\label{fourthline}
\end{eqnarray}

\section{}
\setcounter{equation}{0}
\label{App:E}

Collecting (\ref{sum of first two lines}), (\ref{thirdline}) and (\ref{fourthline})
in Eq.~(\ref{idn4}), 
we obtain the final expression for the diagram $(b)$ in $D$ dimensions
\begin{eqnarray}
I_b = \frac{A(1,1-2\e,1)}{4\e(1+2\e)(1-2\e)^2}
\left\{ \frac{5\e-12\e^2}{[31]^2[12]^{1-\e}} J(\e,2-3\e,1) \right. \no\\  
+ \left. \left[\frac{-1+9\e-17\e^2}{[31]^2[23]^{1-\e}} 
+ \frac{-1+8\e-(23/2)\e^2}{[31][23]^{2-\e}} 
+ \frac{(1-8\e+(23/2)\e^2)[21]}{[31]^2[23]^{2-\e}}  \right] J(1,2-3\e,\e)  \right. \no\\
+ \left. \left[\frac{3\e-(5/2)\e^2}{[31]^2[23]^{1-\e}} 
+ \frac{-(3/2)\e+(5/4)\e^2}{[31][23]^{2-\e}}  
+ \frac{[12](-(3/2)\e+(5/4)\e^2)}{[31]^2[23]^{2-\e}} \right]
(13)_\mu \pd_{\mu}^{(1)}J(1,2-3\e,\e) \right.  \no\\  
+ \left. \frac{1-(5/2)\e^2}{[12]^{1-\e}[31]^2} 
(31)_\mu \pd_{\mu}^{(3)}J(\e,2-3\e,1) \right\}\no\\
\no\\
- \frac{1}{2}\left[ \frac{1}{[31]^2[23]} 
+ \frac{1}{[31][23]^2}  - \frac{[12]}{[31]^2[23]^2} \right] J(1,1,1) \no\\
+ \frac{1}{8} \left\{\frac{-27/2}{[12][23][31]^{2}} 
+ \frac{5}{[12][23]^{2}[31]} +  \frac{1/4}{[12]^{2}[23][31]} 
+ \frac{- 11/4}{[12]^{2}[23]^{2}} + \frac{- 9/4}{[23]^{2}[31]^{2}} 
+  \frac{5/2}{[12]^{2}[31]^{2}}  \right\}\no\\
+ \frac{A(1,1-2\e,1)A(1,1-3\e,1+\e)}{8\e(1+2\e)(1-2\e)^2} 
\left\{\frac{-7/2}{[12]^{1-2\e}[23]^{1-2\e}[31]^{2+2\e}}  
+ \frac{-2}{[12]^{1-2\e}[23]^{2-2\e}[31]^{1+2\e} } \right. \no\\
+ \left. \frac{-1/2}{[12]^{-2\e}[23]^{2-2\e}[31]^{2+2\e}} 
+  \frac{3/2}{[12]^{2-2\e}[23]^{1-2\e}[31]^{1+2\e}} 
+ \frac{3/2}{[12]^{2-2\e}[23]^{2-2\e}[31]^{2\e}} \right.\no\\
+ \left. \frac{3}{[12]^{2-2\e}[23]^{-2\e}[31]^{2+2\e}} + \frac{-6}{[31]^2[12]^{2-2\e}} +  \frac{1}{[31]^2[23]^{2-2\e}} \right\}\no\\
\no\\
+ \frac{A^2(1,1-2\e,1)}{4\e^2(1+2\e)(1-2\e)^2}
\left\{\frac{1-\e}{[12]^{1-\e}[23]^{2-\e}[31]} + \frac{1-\e}{[12]^{1-\e}[23]^{1-\e}[31]^2} - 
\frac{1 - \e}{[12]^{-\e}[23]^{2-\e}[31]^2} \right\} \ .
\end{eqnarray}
Taking the limit $\e \to 0$, the poles in $\e$ disappear and the result is
\begin{eqnarray}
I_b(\e \to 0) =
\frac{1}{8} \left[\frac{7/2}{[12][23][31]^2}\ln\frac{[31]^2}{[12][23]}  
+ \frac{2}{[12][23]^2[31]}\ln\frac{[31]^2}{[12][23]}   \right.\no \\
+ \left. \frac{-3/2}{[12]^{2}[23][31]}\ln\frac{[31]^2}{[12][23]}   
+ \frac{-1/2}{[23]^{2}[31]^2} \ln\frac{[12]^3}{[23][31]^2}  \right.\no\\
+ \left. \frac{3}{[12]^{2}[31]^2}\ln\frac{[23]^3}{[12][31]^2}   
+  \frac{-3/2}{[12]^{2}[23]^{2}} \ln\frac{[31]^2}{[12][23]}   \right]  \no\\
+ \frac{1}{16}\left[\frac{1}{[12][31]^2[23]} + \frac{1}{[12][31][23]^2} 
+ \frac{-1}{[31]^2[23]^{2}} \right]\ln\frac{[12][23]}{[13]^2}\ln\frac{[23]}{[12]}  \no\\
+ \frac{1}{8}\left[\frac{-2}{[31]^2[23]} + \frac{3[12]}{[31]^2[23]^{2}} 
+ \frac{1}{[12][23]^2}   - \frac{1}{[31]^2[12]} 
+ \frac{-4}{[31][23]^2}\right] J(1,1,1) \no\\
- \frac{3}{16}\left[\frac{2}{[31]^2[23]} + \frac{1}{[31][23]^{2}}  
- \frac{[12]}{[31]^2[23]^{2}} \right] (13)_\mu \pd_{\mu}^{(1)} 
\frac{1}{[12]} \ln\frac{[23]}{[12]} \no\\
- \frac{1}{8}\frac{1}{[12][31]^2} (31)_\mu \pd_{\mu}^{(3)}
\frac{1}{[23]}\le\frac{1}{2}\ln\frac{[23][12]}{[13]^2}
\ln\frac{[12]}{[23]} + ([23] + [13] - [12])J(1,1,1) \ri \no\\
+ \frac{1}{8} \left[\frac{7}{[12][23]^{2}[31]} + \frac{3}{[12][23][31]^2} 
+  \frac{-7}{[23]^{2}[31]^2} \right]\ln\frac{[12]}{[23]}\no\\ 
+ \frac{1}{16} \left\{\frac{-13}{[12][23][31]^{2}} 
+ \frac{16}{[12][23]^{2}[31]} +  \frac{-3}{[12]^{2}[23][31]} 
+ \frac{- 9}{[12]^{2}[23]^{2}} + \frac{-7}{[23]^{2}[31]^{2}} 
+  \frac{12}{[12]^{2}[31]^{2}}  \right\} \ . \label{2001}
\end{eqnarray}

\end{appendix}

\end{document}